\pdfoutput=1

\documentclass[conference]{IEEEtran}

\usepackage{microtype}
\usepackage{subcaption}
\usepackage{booktabs} 
\usepackage{arabtex}
\usepackage{multirow}
\usepackage{lipsum}  
\usepackage{algorithm}
\usepackage{algorithmic}
\usepackage{csquotes}
\usepackage[table]{xcolor}
\usepackage{graphicx}
\usepackage{tikz}
\usepackage{amsmath}
\usepackage{amssymb}
\usepackage[bookmarks=false]{hyperref}
\usepackage{listings,multicol}
\usepackage{xspace}
\usepackage{jlcode}
\usepackage{enumitem}
\usepackage{stmaryrd}

\usepackage[capitalize]{cleveref}
\usepackage{bm}

\renewcommand{\vec}{\bm}
\newcommand{\statdp}{{\textsc{StatDP}}\xspace}
\newcommand{\toolname}{{\textsc{Kolahal}}\xspace}

\newcommand{\Alg}{M}

\newcommand{\sk}{M^{\bullet}}
\newcommand{\fsk}[1]{M^{#1}}
\newcommand{\evec}{\vec{\eta}}
\newcommand{\cvec}{\vec{c}}

\newcommand{\Lap}{\ensuremath{\mathsf{Lap}}}
\newcommand{\Exp}{\ensuremath{\mathsf{Exp}}}

\newcommand{\dt}{\mathsf{D}}
\newcommand{\mem}{\mathsf{Mem}}
\newcommand{\var}{\mathsf{Var}}
\newcommand{\expr}{\mathsf{Exp}}

\newcommand{\com}{\mathsf{Com}}

\newcommand{\denot}[1]{\llbracket {#1} \rrbracket}

\newcommand{\pr}{\mathbb{P}}

\DeclareMathOperator*{\argmin}{\arg\!\min}

\renewcommand{\paragraph}[1]{\vspace{.1in}\noindent\textbf{{#1}}}
\newcommand{\loss}{L}

\newcommand{\tuple}[1]{\langle#1\rangle}

\newcommand{\citep}[1]{\cite{#1}}
\newcommand{\citet}[1]{\cite{#1}}

\renewcommand{\leq}{\leqslant}
\renewcommand{\geq}{\geqslant}

\newtheorem{definition}{Definition}[section]

\IEEEoverridecommandlockouts

\setarab
\novocalize
\arabtrue

\begin{document}

\title{Learning Differentially Private Mechanisms}

 \author{\IEEEauthorblockN{Subhajit Roy$^1$}\thanks{$^1$ The work was primarily done during his visit to UW--Madison.}
 \IEEEauthorblockA{\textit{Computer Science and Engineering} \\
 \textit{Indian Institute of Technology Kanpur}\\
 subhajit@iitk.ac.in}
 \and
 \IEEEauthorblockN{Justin Hsu}
 \IEEEauthorblockA{\textit{Department of Computer Sciences} \\
 \textit{University of Wisconsin--Madison}\\
 justhsu@cs.wisc.edu}
 \and
 \IEEEauthorblockN{Aws Albarghouthi$^2$}\thanks{$^2$ Author's name in native alphabet: \RL{'aws albr.gU_ty}}
 \IEEEauthorblockA{\textit{Department of Computer Sciences} \\
 \textit{University of Wisconsin--Madison}\\
 aws@cs.wisc.edu}
 }


\maketitle

\begin{abstract}
Differential privacy is a formal, mathematical definition of data privacy that
has gained traction in academia, industry, and government. The task of correctly
constructing differentially private algorithms is non-trivial, and mistakes have
been made in foundational algorithms. Currently, there is no automated support
for converting an existing, non-private program into a differentially private
version. In this paper, we propose a technique for automatically learning an
accurate and differentially private version of a given non-private program.  We
show how to solve this difficult program synthesis problem via a combination of
techniques: carefully picking representative example inputs, reducing the
problem to continuous optimization, and mapping the results back to symbolic
expressions. We demonstrate that our approach is able to learn foundational
algorithms from the differential privacy literature and significantly
outperforms natural program synthesis baselines.
\end{abstract}


\section{Introduction}

Today, private data from individuals is commonly aggregated in large databases
maintained by companies or governmental organizations.
There is clear value in using this data to learn information about a general
population, but there are also serious privacy concerns---seemingly innocuous
data analyses can unintentionally leak highly sensitive information about
specific
individuals~\citep{DBLP:conf/sp/NarayananS08,DBLP:conf/sp/CalandrinoKNFS11}.

To address these concerns, differential privacy~\cite{DMNS06} is a rigorous,
mathematical definition of data privacy that has attracted a flurry of interest
across academia, industry, and government.
Intuitively, differential privacy defines privacy as a kind of
\emph{sensitivity} property: given a program that takes a private dataset as
input, adding or removing an individual's data should only have a small effect
on the program's output distribution.
While differentially private programs were originally designed by theoreticians
to solve specific statistical tasks, programmers in many areas are now looking
to use differentially-private programs for their application
domain~\cite{erlingsson2014rappor,johnson2018towards}.

However, applying differential privacy is far from easy.
First, to satisfy the mathematical guarantee, differentially private programs
must be carefully crafted to add probabilistic noise at certain key points.
The task of correctly constructing differentially private algorithms is
non-trivial, and mistakes have been discovered in commonly used
algorithms~\cite{lyu2016understanding}.
Second, although there are numerous software packages that provide private
building-blocks and methods for safely combining components while respecting the
privacy guarantee~\cite{mcsherry2009privacy,roy2010airavat}, existing frameworks
for differential privacy are intended for writing new private programs from
scratch: practitioners who have domain-specific code developed without
considering privacy cannot automatically convert their code to differentially
private programs.
Instead, programmers must \emph{reimplement} their code, manually figure out
where to add random noise---while keeping in mind that suboptimal choices may
trivially achieve privacy by adding so much noise as to ruin accuracy---and then
\emph{prove} that the resulting algorithm is indeed differentially private.
For many applications, the programmer burden is simply too high for differential
privacy to be a realistic approach.

\paragraph{Mechanism synthesis problem.}
In this paper, we are interested in \emph{automatically synthesizing a
differentially private program} from an existing, non-private program. More
specifically, we pose the question as follows:  
\begin{displayquote}
  Given a non-private algorithm $M$, can we generate an algorithm $M^\star$ that
  is differentially private and ``accurate'' with respect to $M$?
\end{displayquote}

This problem is difficult for several reasons: 
(1) The space of algorithms is infinite and discrete, and it is not clear how to
search through it to find $M^\star$. Changing the amount of noise added at some
program locations may sharply change the privacy level of the whole algorithm,
while adjusting the noise level at other program locations may have no effect on
privacy.
(2) Given a candidate $M^\star$ that is not differentially private, it may be
difficult to find a counterexample; given a counterexample, it is not obvious
how to make the candidate more private.
(3) While a testing/verification tool is only concerned with proving or
disproving differential privacy, a synthesis method has the additional goal of
finding a mechanism that adds as little noise as possible to achieve the target
privacy level.

To address these problems, we present a set of novel contributions to synthesize
\emph{pure} differentially private (i.e., \mbox{$\epsilon$-differentially} private)
mechanisms.

\paragraph{Our approach.}
First, we restrict the search space for $M^\star$ to variants of $M$ with noise
added to some selected program expressions. These expressions can be selected by
a domain expert and provided to our algorithm as a \textit{mechanism sketch}.
The question now becomes: \emph{how much noise should we add?} In general, the
amount of noise may need to be a function of the algorithm's inputs and
$\epsilon$, the privacy parameter. 

To search the space of noise functions, we employ ideas from \emph{inductive
program synthesis}~\cite{solar2006combinatorial,alur2013syntax}, where candidate
programs are proposed and then refined through counterexamples generated by a
testing tool; we rely on a state-of-the-art tool for detecting violations of
differential privacy, called \statdp~\cite{ding2018detecting}. However, a
na\"ive search strategy runs very slowly, as the testing tool can take a long
time to discover counterexamples to privacy for every candidate we provide. To
make the search process more efficient, we demonstrate how to set up a
continuous optimization problem that allows us to hone in on the most-likely
candidates, eliminating unlikely candidates from the search space without
calling the testing tool. Our optimization problem also guides our search
towards noise parameters such that the privacy guarantee is \emph{tight}, i.e.,
our synthesis procedure aims to find programs that are $\epsilon$-differentially
private, but not $\epsilon'$-differentially private for any smaller $\epsilon'$.
This strategy ensures that there are no obvious places where too much noise is
being added, and enables our tool to find private algorithms that have been
proposed in the privacy literature, as well as \emph{new variants}.

\paragraph{Evaluation.}
We have implemented and applied our approach to synthesize a range of
$\epsilon$-differentially private algorithms. Our results show that (1) our
approach is able to discover sophisticated algorithms from the literature,
including the sparse vector technique (SVT), and (2) our continuous
optimization-guided search improves performance significantly over simpler
approaches.

\paragraph{Contributions.}
We offer the following technical contributions.
\begin{itemize}
  \item We present a sketch-based methodology to construct an
    $\epsilon$-differentially private mechanism $M^{*}$ from a given
    non-private program $M$.
  \item Our technique combines a number of novel ideas: (a) bootstrapping the
    learning process with a number of carefully selected examples, (b) solving
    an approximate continuous optimization variant of the problem to guide an
    enumerative program-synthesis approach, and (c) using the privacy loss to
    help rank the proposed programs, treating it as a proxy for the tightness of
    the privacy guarantee.
  \item We implement our approach in a tool called \toolname,\footnote{%
          ``{\toolname}'' (\vspace{-5pt}\includegraphics[trim={0 7pt 0 0},clip,height=0.95em]{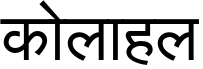}) is a Hindi word meaning \textit{loud noise}.}
    and evaluate it on synthesizing several foundational algorithms from the
    differential privacy literature, e.g., the \emph{sparse vector technique}
    (SVT). We compare our approach to a series of successively stronger baseline
    procedures, demonstrating the importance of our algorithmic choices.
\end{itemize}
Taken together, our work is the first to automatically synthesize complex 
differentially private mechanisms. 

\paragraph{Limitations.}
While there are now many known approaches for verifying differential privacy,
existing methods are too slow to be used in our counterexample-guided synthesis
loop. Hence, our synthesis procedure leverages an efficient, counterexample
generation tool to check if candidates are not private. Note that this tool is
\emph{unsound for verification}: failure to find a counterexample does not prove
differential privacy. Instead, after our tool produces a ranked list of
candidates, each candidate must be analyzed by a sound verifier as a final
check. Secondly, our proposal only attempts to synthesize suitable noise
expressions in mechanism \textit{sketches}---our algorithm does not look to
transform the sketch.  Finally, we have only investigated our method for pure,
$(\epsilon, 0)$-privacy, not variants such as $(\epsilon, \delta)$-privacy, or
R\'enyi differential privacy. Please refer to Section~\ref{sec:discuss} for
details.

\section{Background}

We begin by introducing key definitions and existing algorithmic tools that we
will use in our synthesis procedure. Readers interested in a more thorough
presentation of differential privacy should consult the textbook
by Dwork and Roth~\citet{DR14}.

\subsection{Differential privacy}

Differential privacy~\citep{DMNS06} is a quantitative definition of privacy for
programs that take private information as input, and produce a randomized
answer as output.

\paragraph{Mechanism.}
A \emph{mechanism} (or program) $\Alg$ takes as input a database $d$ of private
information, a \emph{privacy parameter} $\epsilon \in \mathbb{R}_{>0}$, and
potentially other inputs (e.g., queries to be answered on the private database),
then returns a noisy output of type $T$.

\paragraph{Neighboring (or adjacent) databases.}
We shall assume a relation $\Delta$ over pairs of databases. If $(d_1,d_2) \in
\Delta$, then we say that $d_1$ and $d_2$ are \emph{neighboring} (or
\emph{adjacent}) databases. Intuitively, $\Delta$ relates pairs of databases
that differ in a single individual's private data. For instance, $\Delta$ might
relate one database to a neighboring database where one individual's record has
been added, removed, or modified. We assume that this relation is provided as
part of the input specification.

\paragraph{Privacy loss.}
For any pair of databases $(d_1,d_2) \in \Delta$, privacy parameter $\epsilon >
0$, and event $E \subseteq T$, we define the \emph{privacy loss}
$\loss(\Alg,d_1,d_2,\epsilon,E)$ to be: 
\[
\max\left(\frac{\pr[\Alg(d_1,\epsilon) \in E]}{\pr[\Alg(d_2,\epsilon) \in E]}, \frac{\pr[\Alg(d_2,\epsilon) \in E]}{\pr[\Alg(d_1,\epsilon) \in E]}\right)
\]

\paragraph{Pure differential privacy.}
Using our definitions, $\epsilon$-\emph{differential privacy} (DP)---also known
as pure differential privacy~\citep{DMNS06}---is defined as follows: $\Alg$ is
$\epsilon$-DP iff for all $(d_1,d_2) \in \Delta$,  $\epsilon > 0$, and event $E
\subseteq T$, the privacy loss is upper bounded by $e^\epsilon$:
\[ 
    \loss(\Alg,d_1,d_2,\epsilon,E) \leq e^\epsilon
\]
Accordingly, a \emph{counterexample} to $\epsilon$-DP is a tuple
$\tuple{d_1,d_2,\epsilon,E}$ such that $(d_1,d_2)\in\Delta$, $\epsilon>0$, and
$\loss(\Alg,d_1,d_2,\epsilon,E) > e^\epsilon$.

\begin{figure*}
  \begin{subfigure}[b]{0.49\textwidth}
    \begin{align*}
      \var\ x &::= a \mid b \mid \cdots
      \\
      \expr\ e &::= x
      \mid \mathsf{true}
      \mid \mathsf{false}
      \mid e = e'
      \mid e < e'
      \mid e > e'
      \mid \neg e
      \\
              &\mid \mathbb{Z}
              \mid e + e'
              \mid e \cdot e'
              \mid e / e'
              \\
      \com\ c &::= \mathsf{skip}
      \mid x \gets e
      \mid x \gets \Lap(e)(e')
      \\
             &\mid c ; c'
             \mid \mathsf{if}\ e\ \mathsf{then}\ c\ \mathsf{else}\ c'
             \mid \mathsf{while}\ e\ \mathsf{do}\ c
    \end{align*}
    \caption{Language Syntax}
    \label{fig:lang-syntax}
  \end{subfigure}
  \hfill
  \begin{subfigure}[b]{0.49\textwidth}
    \begin{align*}
      \denot{\mathsf{skip}}m &\triangleq \mathsf{unit}(m)
      \\
      \denot{x \gets e}m &\triangleq \mathsf{unit}(m[x \mapsto \denot{e}m])
      \\
      \denot{x \gets e + \Lap(e')}m &\triangleq
      \mathsf{bind}(\mathcal{L}_{\denot{e'}m}(\denot{e}m) , v \mapsto m[x \mapsto v])
      \\
      \denot{c ; c'}m &\triangleq
      \mathsf{bind}(\denot{c}m, m' \mapsto \denot{c'}m')
      \\
      \denot{\mathsf{if}\ e\ \mathsf{then}\ c\ \mathsf{else}\ c'}m &\triangleq
      \begin{cases}
        \denot{c}m &: \text{if } \denot{e}m = \mathsf{true} \\
        \denot{c'}m &: \text{if } \denot{e}m = \mathsf{false}
      \end{cases}
      \\
      \denot{\mathsf{while}\ e\ \mathsf{do}\ c}m &\triangleq
      \lim_{n \to \infty} \denot{(\mathsf{if}\ e\ \mathsf{then}\ c\ \mathsf{else}\ \mathsf{skip})^n}m
    \end{align*}
    \caption{Language Semantics}
    \label{fig:lang-semantics}
  \end{subfigure}

  \caption{Programming Language}
  \label{fig:language}
\end{figure*}

\subsection{Programming language: syntax and semantics}

We will work with an imperative language with a random sampling command,
described in \cref{fig:language}. The syntax and semantics are largely standard;
readers who are more interested in the synthesis procedure can safely skip
ahead.

The syntax is summarized in \cref{fig:lang-syntax}. Here, $\var$ is a countable
set of program variables, and $\expr$ is a set of program expressions.
Expressions may be boolean- or integer-valued; we implicitly assume that all
expressions are well-typed. Besides the usual commands in an imperative
language, there are two constructs that might be less familiar: $\mathsf{skip}$
is the do-nothing command, and $x \gets e + \Lap(e')$ draws a random sample from
the Laplace distribution with mean $e$ and scale $e'$, and stores the result in
variable $x$; this commands is one of the building blocks for differentially
private algorithms. Throughout, we will use the abbreviation:
\[
  \mathsf{if}\ e\ \mathsf{then}\ c \triangleq
  \mathsf{if}\ e\ \mathsf{then}\ c\ \mathsf{else}\ \mathsf{skip}
\]

Our semantics for commands, summarized in \cref{fig:lang-semantics}, is also
largely standard. The program state is modeled by a memory $m \in \mem$, where
$\mem$ is the set of maps from variables in $\var$ to values. We model commands
$c$ as functions \mbox{$\denot{c} : \mem \to \dt(\mem)$}, taking an input memory to a
distribution over output memories, where a distribution $\mu \in \dt(\mem)$ is
a map $\mu : \mem \to [0, 1]$ such that $\mu$ has countable support (i.e., there
are countably many $m \in \mem$ such that $\mu(m) \neq 0$) and the weights in
$\mu$ sum up to $1$.

The formal semantics in \cref{fig:lang-semantics} uses two operations on
distributions. First, given $a \in A$, the \emph{Dirac distribution}
$\mathsf{unit}(a) \in \dt(A)$ is defined via
\[
\begin{cases}
  \mathsf{unit}(a)(a') \triangleq 1 &: a = a' \\
  \mathsf{unit}(a)(a') \triangleq 0 &: a \neq a' .
\end{cases}
\]
That is, $\mathsf{unit}$ is simply the point mass distribution. Second, given
$\mu \in \dt(A)$ and $f : A \to \dt(B)$, the \emph{distribution bind}
$\mathsf{bind}(\mu, f) \in \dt(B)$ is defined via
\[
  \mathsf{bind}(\mu, f)(b) \triangleq \sum_{a \in A} \mu(a) \cdot f(a)(b) .
\]
Intuitively, $\mathsf{bind}$ sequences a distribution together with a
continuation, leading to a single distribution on outputs.

We make two brief remarks about the semantics. First, the semantics of sampling
involves the (discrete) Laplace distribution $\mathcal{L}_b(z)$, where $z \in
\mathbb{Z}$ is the \emph{mean} of the distribution and $b \in \mathbb{R}$ is a
positive \emph{scale} parameter. The distribution $\mathcal{L}_b(z) \in
\dt(\mathbb{Z})$ is defined as follows:\footnote{%
  The standard Laplace distribution has support over the real numbers; we take
the discretized version to simplify the technical development.}
\[
  \mathcal{L}_b(z)(z') \triangleq \frac{\exp(- |z - z'| / b)}{\sum_{y \in
\mathbb{Z}} \exp(- |z - y| / b)}
\]
Intuitively, the scale parameter $b$ controls how broadly spread the
distribution is---larger values of $b$ lead to a more spread out distribution.
Second, the semantics of loops is well-defined provided that the loop terminates
with probability $1$ on any input. Since this property holds for all private
(and non-private) programs of interest, we will assume this throughout the
paper.

\subsection{Testing for differential privacy}

At a high level, our synthesis procedure iteratively tries different settings of
the unknown noise parameters. Initial candidates will almost certainly fail to
satisfy differential privacy. To make progress, our procedure leverages
\statdp~\citep{ding2018detecting}, a counterexample generation tool for
differential privacy. Given a mechanism $\Alg$ and a target privacy level
$\epsilon_0$, \statdp constructs a set of candidate counterexamples $\{(d_1,
d_2, E)\}$ that may witness a differential privacy violation; here $d_1$, $d_2$
are adjacent databases and $E$ is a subset of outputs that is much more likely
on one database than on the other.

Since our synthesis procedure leverages more specialized information provided by
\statdp, we briefly describe how \statdp operates. \statdp uses a set of
patterns to generate test databases $(d_1, d_2)$, and a set of heuristics to
construct test events $E$. For each test, \statdp estimates the probabilities
$\tilde{\rho_1}, \tilde{\rho_2}$ of the output being in $E$ starting from inputs
$d_1$ and $d_2$ respectively, by repeatedly running the given mechanism. Then,
it runs a \emph{hypothesis test} (Fisher's exact test) to decide how likely the
true probabilities $\rho_1$ and $\rho_2$ are to satisfy the guarantee that the
program is differential private at some \textit{test} $\epsilon$ values in the
neighborhood of target level of privacy $\epsilon_0$. For instance, if
$\epsilon_0=0.5$, \statdp will test whether the mechanism is $\epsilon$-DP for
$\epsilon \in \{0.4, 0.5, 0.6\}$. If the hypothesis test indicates that $\rho_1$
and $\rho_2$ are highly unlikely to satisfy the privacy guarantee, then \statdp
returns $(d_1, d_2, E)$ as a candidate counterexample. Along with this tuple,
\statdp also reports the $p$-\emph{value} of the statistical test, a number in
$[0, 1]$ measuring the confidence: a small $p$-value indicates that the
candidate is likely to be a true counterexample to $\epsilon$-differential
privacy, while a large $p$-value indicates that the candidate is unlikely to be
a true counterexample.

\begin{figure}[t]
    \centering
        \includegraphics[scale=0.5]{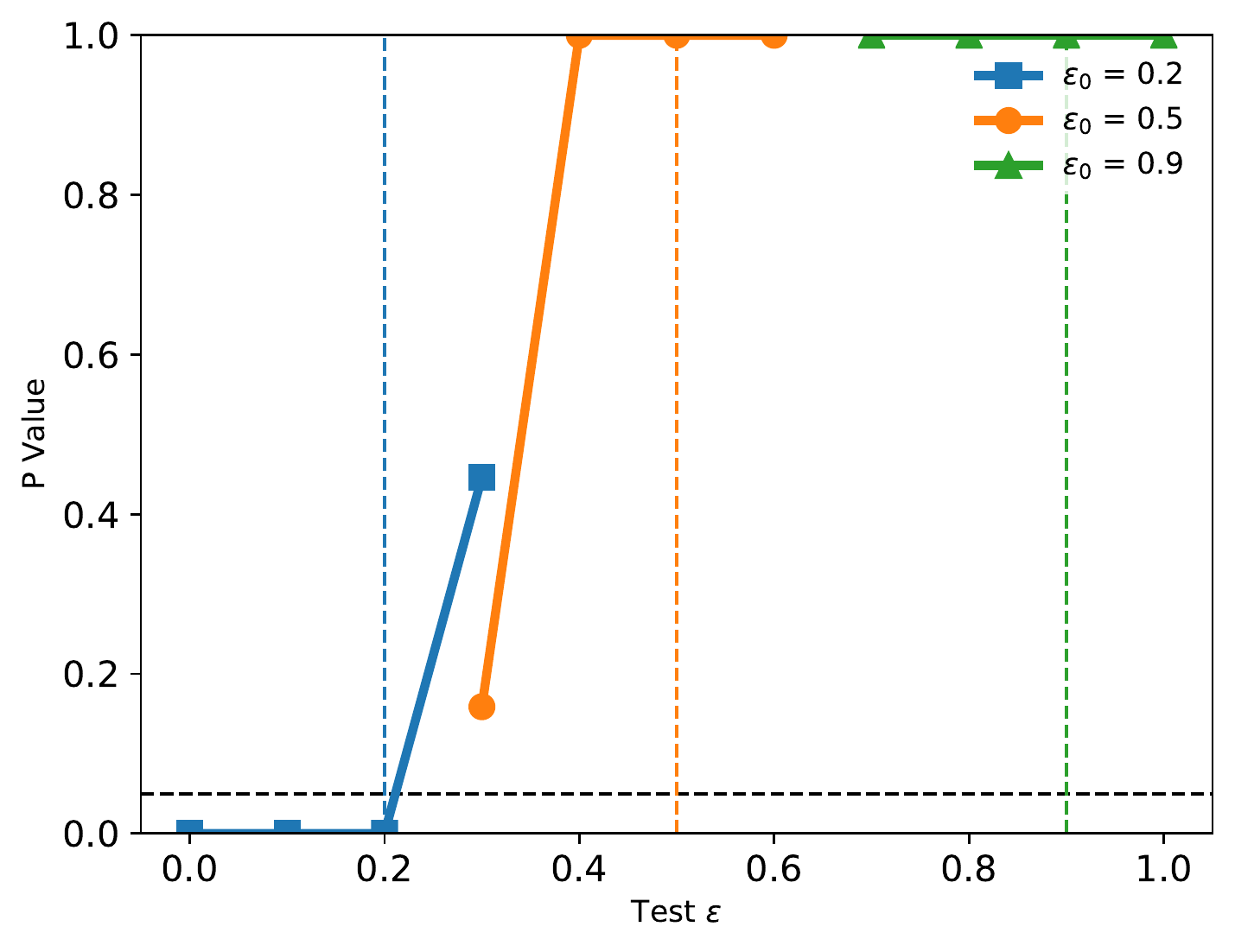}
        \caption{\label{fig:dptest}\statdp at 0.2, 0.5, and 0.9}
\end{figure}

\Cref{fig:dptest} shows the $p$-values produced by a run of \statdp on a
particular $0.5$-differentially private mechanism $M$, checking against target
privacy levels $\epsilon_0$ of 0.2, 0.5 and 0.9. Since $M$ is
$0.5$-differentially private, it is automatically also $0.9$-differentially
private; however, $M$ is not necessarily $0.2$-differentially private. The
candidate counterexamples returned by \statdp reflect this description. For
small test $\epsilon$, \statdp produces a candidate counterexample with a
$p$-value close to zero, meaning that the candidate is highly likely to be a
counterexample to $\epsilon$-DP. For large test $\epsilon$, \statdp still
produces a candidate counterexample but with a $p$-value close to one---this
indicates that the candidate is unlikely to be a true counterexample to
$\epsilon$-DP. Near the true privacy level ($\epsilon \in [0.2, 0.4]$), the
reported $p$-value is in the middle of the range $[0, 1]$---\statdp is uncertain
whether it has found a true counterexample or not. 

Our synthesis approach crucially relies on the $p$-value reported by \statdp to
judge the difficulty of the candidate counterexamples. Difficult cases---ones
where $p$ is in the middle of the range $[0, 1]$---are used as examples to
improve the privacy of the synthesized mechanism. We will discuss this heuristic
further in \cref{sec:algorithm}, but intuitively, challenging counterexamples to
a mechanism represent pairs of inputs and sets of outputs where the privacy
guarantee is nearly \emph{tight}: lowering the target privacy level $\epsilon_0$
would cause these challenging counterexamples to turn into true counterexamples
to $\epsilon_0$-privacy. As a result, challenging counterexamples witness the
fact that a particular mechanism is \emph{almost exactly} \mbox{$\epsilon_0$-private},
and not more private than desired. Mechanisms without challenging
counterexamples are likely more private than necessary, adding more noise than
is needed.

We survey other approaches for testing and verification of differential privacy
in related work (\cref{sec:rw}).

\section{Overview}\label{sec:problem}

\subsection{An illustrative example}

\begin{figure*}
  \begin{subfigure}[b]{0.29\textwidth}
    \[
      \begin{array}{l}
        \mathsf{aboveT}(d, \vec{q}, T): \\
        \quad i \gets 1; \\
        \quad done \gets \mathsf{false}; \\
        \quad t \gets T \vphantom{\boxed{\eta_1}}; \\
        \quad \vec{a} \gets \vec{q}(d) \vphantom{\boxed{\eta_1}}; \\
        \quad \mathsf{while}\ i \leq |\vec{q}| \land \lnot done\ \mathsf{do}: \\
        \quad\quad \mathsf{if}\ a_i > t\ \mathsf{then} \\
        \quad\quad\quad done \gets \mathsf{true}; \\
        \quad\quad i \gets i + 1; \\
        \quad \mathsf{if}\ done \\
        \quad\quad ans \gets i - 1; \\
        \quad \mathsf{else} \\
        \quad\quad ans \gets 0; \\
        \quad out \gets ans; \\
        \quad \mathsf{return}~out;
      \end{array}
    \]
  \caption{Original program}
  \label{fig:ex-aboveT-orig}
  \end{subfigure}
  \hfill
  \begin{subfigure}[b]{0.34\textwidth}
    \[
      \begin{array}{l}
        \mathsf{aboveT}^\bullet(d, \vec{q}, T, \epsilon): \\
        \quad i \gets 1; \\
        \quad done \gets \mathsf{false}; \\
        \quad t \gets T + \Lap(\boxed{\eta_1}); \\
        \quad \vec{a} \gets \vec{q}(d) + \overrightarrow{\Lap}(\boxed{\eta_2}); \\
        \quad \mathsf{while}\ i \leq |\vec{q}| \land \lnot done\ \mathsf{do}: \\
        \quad\quad \mathsf{if}\ a_i > t\ \mathsf{then} \\
        \quad\quad\quad done \gets \mathsf{true}; \\
        \quad\quad i \gets i + 1; \\
        \quad \mathsf{if}\ done \\
        \quad\quad ans \gets i - 1; \\
        \quad \mathsf{else} \\
        \quad\quad ans \gets 0; \\
        \quad out \gets ans + \Lap(\boxed{\eta_3}); \\
        \quad \mathsf{return}~out;
      \end{array}
    \]
  \caption{Input: Sketch program}
  \label{fig:ex-aboveT-sketch}
  \end{subfigure}
  \hfill
  \begin{subfigure}[b]{0.35\textwidth}
    \[
      \begin{array}{l}
        \mathsf{aboveT}^\star(d, \vec{q}, T, \epsilon): \\
        \quad i \gets 1; \\
        \quad done \gets \mathsf{false}; \\
        \quad t \gets T + \Lap(2/\epsilon) \vphantom{\boxed{\eta_1}}; \\
        \quad \vec{a} \gets \vec{q}(d) + \overrightarrow{\Lap}(4/\epsilon) \vphantom{\boxed{\eta_1}}; \\
        \quad \mathsf{while}\ i \leq |\vec{q}| \land \lnot done\ \mathsf{do}: \\
        \quad\quad \mathsf{if}\ a_i > t\ \mathsf{then} \\
        \quad\quad\quad done \gets \mathsf{true}; \\
        \quad\quad i \gets i + 1; \\
        \quad \mathsf{if}\ done \\
        \quad\quad ans \gets i - 1; \\
        \quad \mathsf{else} \\
        \quad\quad ans \gets 0; \\
        \quad out \gets ans + \Lap(\bot) \vphantom{\boxed{\eta_1}}; \\
        \quad \mathsf{return}~out;
      \end{array}
    \]
  \caption{Output: Private program}
  \label{fig:ex-aboveT-priv}
  \end{subfigure}
  \caption{From a non-private sketch to a private program.}
  \label{fig:ex-aboveT}
\end{figure*}

To better illustrate the goal of our algorithm, consider the three programs in
\cref{fig:ex-aboveT}. The first program, \cref{fig:ex-aboveT-orig}, takes in a
database $d$ of private information, a list of numeric queries $\vec{q}$, and a
numeric threshold $T$, and checks if any query has answer~(\textit{ans}) above the threshold~($T$)
when evaluated on the database. We suppose that this program is written by
someone without taking privacy into account---the program adds no noise, and it
does not satisfy differential privacy.

Our aim is to \emph{automatically convert} this algorithm into a
differentially-private version. To start with, we create a \emph{sketch} of the
program, \cref{fig:ex-aboveT-sketch}, where the desired privacy level $\epsilon$
is given as an input, and where certain assignment instructions are marked to
add noise from the \emph{Laplace distribution}, a standard numerical
distribution used in differential privacy. These program locations can be identified
by a domain expert, or they can simply be taken to be every assignment
instruction. However, selecting \emph{where} to add noise is not enough---the
Laplace distribution requires a \emph{scale parameter} describing the amount of
noise that is added. Setting parameters incorrectly will lead to an program that
is not $\epsilon$-differentially private, or a program that adds too much noise.
These unknown parameters are indicated by boxes in \cref{fig:ex-aboveT-sketch};
note that the second location adds a \emph{vector} of independent Laplace draws
with the same, unknown scale parameter to the vector $\vec{q}(d)$.

Then, our algorithm searches for a \emph{symbolic expression} for each of the
boxed locations so that the resulting program is $\epsilon$-differentially
private. One of the possible solutions from our algorithm is shown in \cref{fig:ex-aboveT-priv}. Note that
our procedure sets the scale parameter of the last location $\eta_3$ to $\bot$,
indicating that this location does not need any noise. The completed
program---generated by our approach---is known as the \emph{Above Threshold}
algorithm in the differential privacy literature~\citep{DR14}. Along with this
version, our algorithm also finds other differentially private variants; e.g., a
version where the first two noise locations add noise with scale $3/\epsilon$.

This example shows some of the challenges in converting non-private programs to
private versions. First, the target noise parameters are not just constants:
they are often \emph{symbolic expressions}. For instance, the parameters for
$\mathsf{aboveT}$ depend on the symbolic privacy parameter $\epsilon$, and in
general, parameters can depend on other arguments (e.g., the size of an input
list). Second, not all locations need noise. For example, a different way of
completing $\mathsf{aboveT}$ would also add noise at the last location, when
assigning $ans$. While this version would also satisfy $\epsilon$-differential
privacy, it is inferior to the version in \cref{fig:ex-aboveT-priv} since it
adds noise unnecessarily.

\subsection{The Mechanism-Synthesis problem}

We formalize the overall problem as a synthesis problem.

\paragraph{Mechanism sketch.}
A \emph{mechanism sketch} $\sk$ is an incomplete program with \emph{holes} at
certain program locations; these holes represent unknown noise parameters that
must be synthesized in order to produce a differentially private program. The
sketch also specifies (i) the name of the input variable holding the private
input, (ii) an adjacency relation on private inputs, and (iii) the names of
non-private, auxiliary inputs, which we call \emph{arguments}. For
$\mathsf{aboveT}$, for example, the private input is $d$, while the list of
queries ($\vec{q}$) and the threshold ($T$) appear as the arguments. Pairs of
databases $d, d'$ where $\vec{q}(d)$ and $\vec{q}(d')$ differ by at most $1$ in
each coordinate are adjacent.

To complete the mechanism, we need to replace each hole with a well-typed
expression constructed from the program inputs. We will use $\evec$ to denote a
vector of expressions, and $\fsk{\evec}$ to denote the completion of sketch
$\sk$ with the expressions in $\evec$. Recall \cref{fig:ex-aboveT-sketch} for an
example of a sketch, where $\evec$ is a vector of length 3.

Given a sketch $\sk$, there are infinitely many ways of completing the sketch.
To make this problem more tractable, we restrict the space of expressions using
a finite grammar $G$ whose elements we can enumerate.

\paragraph{A proxy for accuracy.}
Even after restricting the possible expressions, there are often many possible
solutions giving $\epsilon$-DP mechanisms; for instance, it is usually quite
easy to construct a $\epsilon$-DP mechanism by selecting an enormous noise
parameter at every location. However, a solution that adds too much noise is
less accurate and less useful. Since directly estimating a concrete mechanism's
accuracy is challenging---for instance, it is often not clear how accuracy
should be defined, and inputs that lead to inaccurate results may be hard to
find---we will use \emph{privacy loss} as a proxy for accuracy.  Intuitively, we
want to find a mechanism with a privacy loss that is \emph{exactly} equal to
$e^\epsilon$---such a mechanism satisfies $\epsilon$-DP tightly, in the sense
that it adds just enough noise to satisfy differential privacy for the target
level of $\epsilon$. Mechanisms with privacy loss below $e^\epsilon$ satisfy
$\epsilon'$-DP for $\epsilon' < \epsilon$, a stronger guarantee that requires
adding more noise.

We will say that an $\epsilon$-DP mechanism $\Alg$ \emph{dominates} $\Alg'$,
denoted $\Alg' \sqsubseteq \Alg$, iff for all pairs of databases $(d_1,d_2) \in
\Delta$, privacy parameters $\epsilon$, and events $E$, we have
\[
  e^\epsilon \geq \loss(\Alg,d_1,d_2,\epsilon,E) \geq \loss(\Alg',d_1,d_2,\epsilon,E) .
\]
Note that $\sqsubseteq$ is a partial order, as some mechanisms are incomparable.
We are now ready to define our mechanism synthesis problem.

\begin{definition}[Problem statement]\label{def:prob}
Given a sketch $\sk$, an optimal solution to the \emph{synthesis problem} is a
vector of expressions $\evec$ such that given any privacy parameter $\epsilon >
0$, the completed mechanism $\fsk{\evec}$ is  
\begin{enumerate}[topsep=2pt,itemsep=2pt,leftmargin=*]
    \item $\epsilon$-differentially private, and 
    \item a maximal mechanism per the ordering $\sqsubseteq$.
\end{enumerate}
\end{definition}

The first point ensures that the mechanism is differentially private, while the
second point ensures that we cannot lower the amount of noise while still
meeting the target differential privacy guarantee. While it is not feasible to
certify that the second condition holds, the ordering helps guide our search
towards more accurate private mechanisms.

\section{Mechanism Synthesis Algorithm}
\label{sec:algorithm}

In this section, we present our technique for synthesizing optimal mechanisms.
A simplistic strategy to synthesizing the mechanism would be to leverage an
enumerative synthesis strategy that proposes expressions from a grammar $G$, and
uses a verifier to accept or reject solutions. However, there are several
interrelated challenges:

\paragraph{Challenge 1: Infinitely many inputs and output events.}
Even with a finite grammar $G$, finding an optimal solution $\fsk{\evec}$
requires showing that it is differentially private and more accurate than (or
incomparable to) all other mechanisms $\fsk{\evec'}$ that are differentially
private. This is challenging due to the universal quantifier over neighboring
databases, privacy parameter, and events. \emph{We solve this challenge by
approximating the universal quantifier with a finite number of carefully chosen
neighboring databases, program arguments, and events.}

\paragraph{Challenge 2: Expensive search over noise expression.}
Even if we have fixed an input, checking every possible expression to see if it
satisfies $\epsilon$-DP on those inputs is an expensive hypothesis testing
process. \emph{To reduce the cost, we adopt a two-phase approach: we first
  approximate the search problem using a fast continuous optimization procedure
where we solve for constant instantiations of the noise values, then search for
symbolic expressions that are close to these constants.}

\paragraph{Challenge 3: Achieving privacy while limiting noise.}
For a given program sketch and a given level of $\epsilon$, there are many
possible ways of adding noise so that the program is $\epsilon$-DP; for
instance, privacy can usually be ensured by adding a large amount of noise at
every location. However, adding too much noise reduces the accuracy of the
algorithm. \emph{To guide our search towards better private algorithms, our
optimization objective takes the tightness of the privacy guarantee and the
sparsity of the noise parameters into account.}

\newcommand{\concparams}{\mathsf{fixParams}}
\newcommand{\generalize}{\mathsf{generalizeExpr}}
\newcommand{\selex}{\mathsf{selectExamples}}
\newcommand{\focus}{\mathsf{getNoiseRegion}}
\newcommand{\search}{\mathsf{findExpr}}
\newcommand{\synth}{\mathsf{synth}}
\newcommand{\exs}{\mathit{Ex}}
\newcommand{\test}{\mathsf{test}}
\newcommand{\conf}{\emph{Conf}}
\newcommand{\ubs}{\emph{Ubs}}
\newcommand{\dir}{\emph{Dir}}
\newcommand{\nbhd}{\emph{Nbhd}}
\newcommand{\conc}{\gamma}

\subsection{High-level mechanism synthesis algorithm}

Our high-level algorithm proceeds in three steps, as shown in
Figure~\ref{fig:synth}.
First, we fix all of the mechanism's inputs ($\gamma$) besides the private
database to concrete values using $\concparams$; these initial arguments can be
drawn from a fixed set of default values, or supplied by the user.
The resulting sketch $\conc(\sk)$ has only one input---the private
database---but it still has holes for unknown noise expressions.
Since we are searching for target expressions that are formed from the
mechanism's non-private inputs, concretizing these inputs means that we can
search for concrete instantiations---real numbers---for each hole.
This transforms the more challenging expression search problem into a simpler
(but still challenging) numerical search problem.

Next, we generate a set of \emph{challenging examples} for $\conc(\sk)$ using
$\selex$, such that ensuring $\epsilon$-differential privacy for these examples
is likely to ensure $\epsilon$-differential privacy \textit{for all inputs}. An
example is a pair of neighboring databases and an output event.
These challenging examples are generated by repeatedly concretizing the program
holes with real numbers, and searching for counterexamples $\exs$ for
$\conc(\sk)$ to $\epsilon$-DP on \statdp.
Then, for the challenging examples, our algorithm searches for a setting of hole
completions so that $\epsilon$-DP holds with the smallest loss to accuracy.
To do so, we construct an optimization problem over possible concrete values of
the noise expressions that aims to make the least-private example in $\exs$ as
tight as possible.
By approximately solving this optimization problem (in the procedure $\focus$), we can extract a region ($R$) of
possible noise values ensuring $\epsilon$-differential privacy on the
given examples.

Finally, we employ an \emph{enumerative synthesis loop} ($\search$) to
\emph{generalize} the concrete instantiations of the noise parameters into a
ranked list of candidate completions $\evec_1,\ldots,\evec_n$---vectors of
symbolic expressions---that give $\epsilon$-differentially private mechanisms
for the arguments $\conc$ as well as other arguments~$\conc'$.
Expressions whose concrete values on $\conc$ do not belong to the region
$R$ can be pruned immediately, without testing differential privacy.
This allows us to \textit{focus} the expensive task of testing differential
privacy to a few selected symbolic expressions whose concretizations lie in this noise
region $R$.

We now describe each step in detail. We will use the program $\mathsf{aboveT}$
from \cref{sec:problem} as our running example; \cref{fig:ex-aboveT-sketch}
shows a possible sketch, and \cref{fig:ex-aboveT-priv} shows a completion of the
program satisfying differential privacy.

\begin{figure}[t]
  \[
    \begin{array}{l}
      \synth(\sk,G):\\
      \quad \conc \gets \concparams(\sk) \\
      \quad \exs \gets \selex(\conc(\sk))\\
      \quad R \gets \focus(\conc(\sk),\exs)\\
      \quad \evec_1,\ldots,\evec_n \gets \search(\sk,G,R,\exs)
    \end{array}
  \]
  \caption{High-level algorithm $\synth$}\label{fig:synth}
\end{figure}

\subsection{Fixing arguments and selecting examples}
Differentially-private mechanisms often take inputs besides the private
database; we call such auxiliary inputs \emph{arguments}. We first fix all
arguments to be some initial values $\conc$, so that the only input of the
sketch $\gamma(\sk)$ is the private database. We assume that sketches take the
target level of privacy $\epsilon$ as a parameter, so $\conc$ also fixes this
variable to some concrete number $\gamma(\epsilon)$. (Our tool, described in
\cref{sec:eval}, actually uses multiple settings of $\conc$ that helps when 
synthesizing symbolic expressions;
for simplicity, we will present our core algorithm using a single setting of
$\conc$.)

There are several reasonable ways to
choose $\conc$. If representative inputs are available---perhaps from the
original, non-private program---these inputs are natural choices for $\conc$.
Otherwise, we can leverage tools capable of producing counterexamples to
differential privacy; these tools produce settings for all inputs to the
program, including the non-private arguments. For instance, the \statdp
tool~\citep{ding2018detecting} uses a combination of symbolic execution and
statistical testing to find counterexample inputs. 

Next, we find a set of examples to bootstrap our synthesis algorithm. More
precisely, an example is a tuple $\tuple{d_1,d_2,E}$ consisting of a pair of
neighboring databases and a set of outputs (an \emph{event}). Intuitively, we
use the examples to quickly screen out choices for the noise scales that
\emph{don't} lead to differentially-private programs---if $M$ is differentially
private, it must have similar probabilities of producing an output in $E$ from
inputs $d_1$ and $d_2$; this local property can be quickly checked without
running the whole testing tool. However, not all examples are equally useful.
Mechanisms may need only a low level of noise to satisfy the privacy condition
at easier examples, but may need higher levels of noise to satisfy the privacy
condition at more difficult examples. Since the differential privacy property
quantifies over all pairs of adjacent databases, we need to find
\emph{challenging} examples that maximizes the privacy loss.

To discover such examples, we leverage \statdp to generate pairs of databases
and output events. Since \statdp requires a complete mechanism as input, rather
than just a sketch, we first complete $\conc(\sk)$ by filling holes with
concrete noise values (i.e., real numbers). To search the space of noise, a
na\"ive grid search over the space of noise values will generate many values of
$\cvec$, but calling \statdp for each $\cvec$ is expensive. Thus instead of
performing a full grid search for noise values in each dimension, we choose a
set $\dir$ of predefined directions and perform a line search along each
direction. The directions are essentially chosen to contain a basis of the
noise space, and hence, together, are likely to create a good representation of
the space. For example, with $n = 2$ noise locations, we choose the vectors as
$\{ \langle1,1\rangle, \langle1,0\rangle, \langle0,1\rangle \}$.

Finally, we apply \statdp on the completed sketch to check if the mechanism is
$\gamma(\epsilon)$-differentially private. If the tester judges the mechanism to
not be differentially private, it returns a counterexample and a $p$-value,
indicating the degree of confidence that the counterexample is a true
counterexample. If the $p$-value of the discovered counterexample is in the
\emph{zone of confusion}, $\conf \subset [0, 1]$---indicating that the tool had
a hard time proving or disproving privacy---then we consider the counterexample
to be challenging and we keep it. Summing up, the set $\selex(\sk)$ is defined as:
\begin{align*}
  \{ \tuple{d_1,d_2,E} \mid{} & (\tuple{d_1,d_2,E}, p) = \test(\gamma(\fsk{\cvec}), \gamma(\epsilon)), \\
                              & p \in \conf, \cvec \in \dir \} 
\end{align*}

\paragraph{Running example: Above Threshold.}
The sketch $\mathsf{aboveT}^\bullet$ from \cref{fig:ex-aboveT-sketch} has three
arguments (non-private inputs): the queries $\vec{q}$, the threshold $T$, and
the target privacy level $\epsilon$. One possible setting of the arguments is:
\[
  \conc = \{ \vec{q} \mapsto (q_1, \dots, q_5), T \mapsto 2, \epsilon \mapsto 0.5 \}
\]
where we abbreviate the queries $q_1, \dots, q_5$. For this setting, our tool
identifies several challenging counterexamples $\tuple{d_1, d_2, E}$, including
$\tuple{r, t, \{ 3 \}}$ and $\tuple{s, t, \{ 3 \}}$, where:
\begin{align*}
  q_1(r), \dots, q_5(r) &= 0, 0, 0, 0, 0 \\
  q_1(s), \dots, q_5(s) &= 2, 2, 0, 0, 0 \\
  q_1(t), \dots, q_5(t) &= 1, 1, 1, 1, 1 .
\end{align*}

\subsection{Reducing to a continuous optimization problem}
After identifying a set of challenging examples $\exs$, we try to find a
concrete instantiation $\cvec$ of $\conc(\fsk{\cvec})$ that (1) is
differentially private at every example in $\exs$, and (2) maximizes the privacy
loss while remaining $\epsilon$-DP; this second criteria biases the search
towards noise parameters that achieve a tight privacy guarantee. Note that this
task is in line with our problem statement (\cref{def:prob}), except that, (a) we
work with a \emph{finite set} of examples rather than all pairs of neighboring
databases and output events, and (b) we complete $\sk$ with \emph{concrete
numbers}, not symbolic expressions.

To find a concrete noise value $\cvec$, we set up the following optimization problem:
\[ \argmin_{\cvec} \left|\left(\max_{\tuple{d_1,d_2,E}\in \exs} \loss(\conc(\fsk{\cvec}), d_1, d_2,\conc(\epsilon),E)\right) - e^{\conc(\epsilon)} \right| \]
This optimization objective looks for a concrete completion~$\cvec$ that makes
the privacy loss of the worst example in $\exs$ as close as possible to the
target privacy loss $e^{\conc({\epsilon})}$. Intuitively, taking the
  objective function to be the absolute difference between the target privacy
  level and the privacy loss at the challenging examples induced by a particular
  noise vector $\cvec$ penalizes concrete noise values $\cvec$ that add too much
  noise or too little noise. If the privacy loss at $\cvec$ is above
  $e^{\conc({\epsilon})}$ then the concretized mechanism should be rejected
  since it fails to satisfy $\conc({\epsilon})$-privacy. If the privacy losss at
  $\cvec$ is below $e^{\conc({\epsilon})}$, then the concretized mechanism
  satisfies a differential privacy guarantee that is \emph{stronger} than the
target guarantee of $\conc({\epsilon})$-privacy; this is not preferred as it adds \emph{more} noise than necessary, typically leading to a less accurate mechanism.

To bias the search towards solutions that add noise at fewer locations, we also
regularize the objective with the $L_0$ (sparsity) norm of $\cvec$; recall that
this norm counts the number of non-zero entries. The final optimization function
is as follows:
\[
    \scalebox{0.9}{$\displaystyle\argmin_{\cvec} \left|\left(\max_{\tuple{d_1,d_2,E}\in \exs}
    \loss(\conc(\fsk{\cvec}), d_1, d_2,\conc(\epsilon),E)\right) - e^{\conc(\epsilon)}
  \right| + \lambda \|\cvec\|_0 $}
\]
where $\lambda\in\mathbb{R}$ is a regularization parameter.

Unfortunately, the optimization problem is not easy to solve---the objective is
not convex or differentiable, and the loss function is expensive to evaluate
even approximately, as it involves computing the probability of invocations of
mechanisms returning certain events.

To approximately solve this optimization problem, we employ an evolutionary
algorithm called \emph{differential evolution}~\citep{Kenneth2005}. The
algorithm maintains a set of candidate solutions, referred to as a
\textit{population}. In each iteration, every candidate in the current
population is moved around using a simple randomized heuristic; the candidate's
new position is retained if it reduces the objective function (referred to as
the \textit{fitness function}). As a result, the population tends to stabilize
in the low loss regions. As the algorithm does not require any other information
about the fitness function except its value at a given candidate, it can be used
for noisy problems that are not differentiable, and not even continuous. Though
differential evolution does not guarantee convergence to the optimal, it is
useful when we are interested in estimating a ``near optimal" \textit{region} of
solutions---captured by the final population.

After multiple rounds of evolutionary refinement, the candidate noise vectors
tends to stabilize, providing us with a set of instantiations that minimize the
objective; we call these candidates the \emph{noise region} $R$.

\begin{figure*}[t]
    \begin{center}
        \includegraphics[scale=0.65, trim=45 20 10 10,clip]{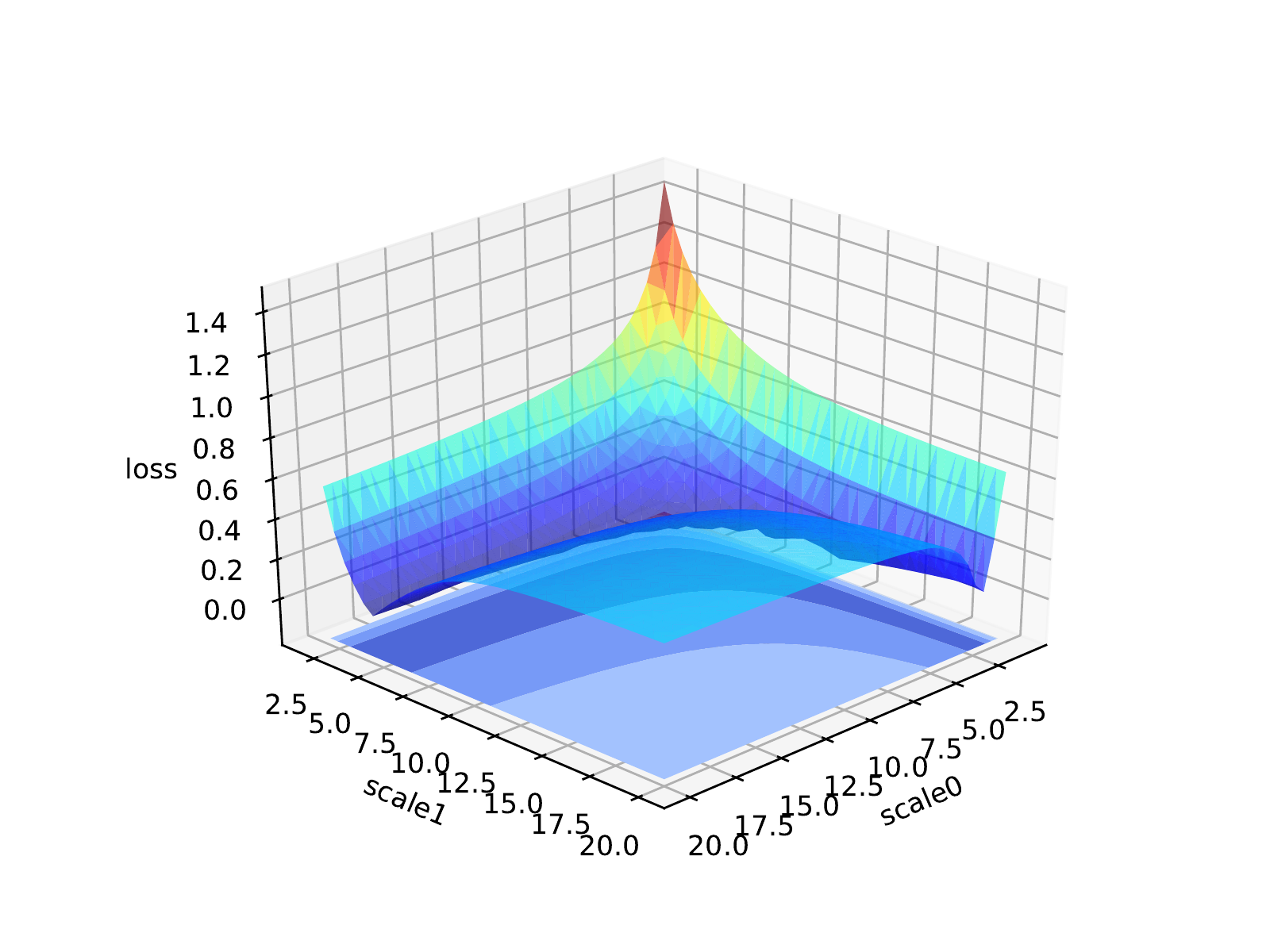}%
        \hspace*{-20pt}\includegraphics[scale=0.67, trim=50 40 10 10,clip]{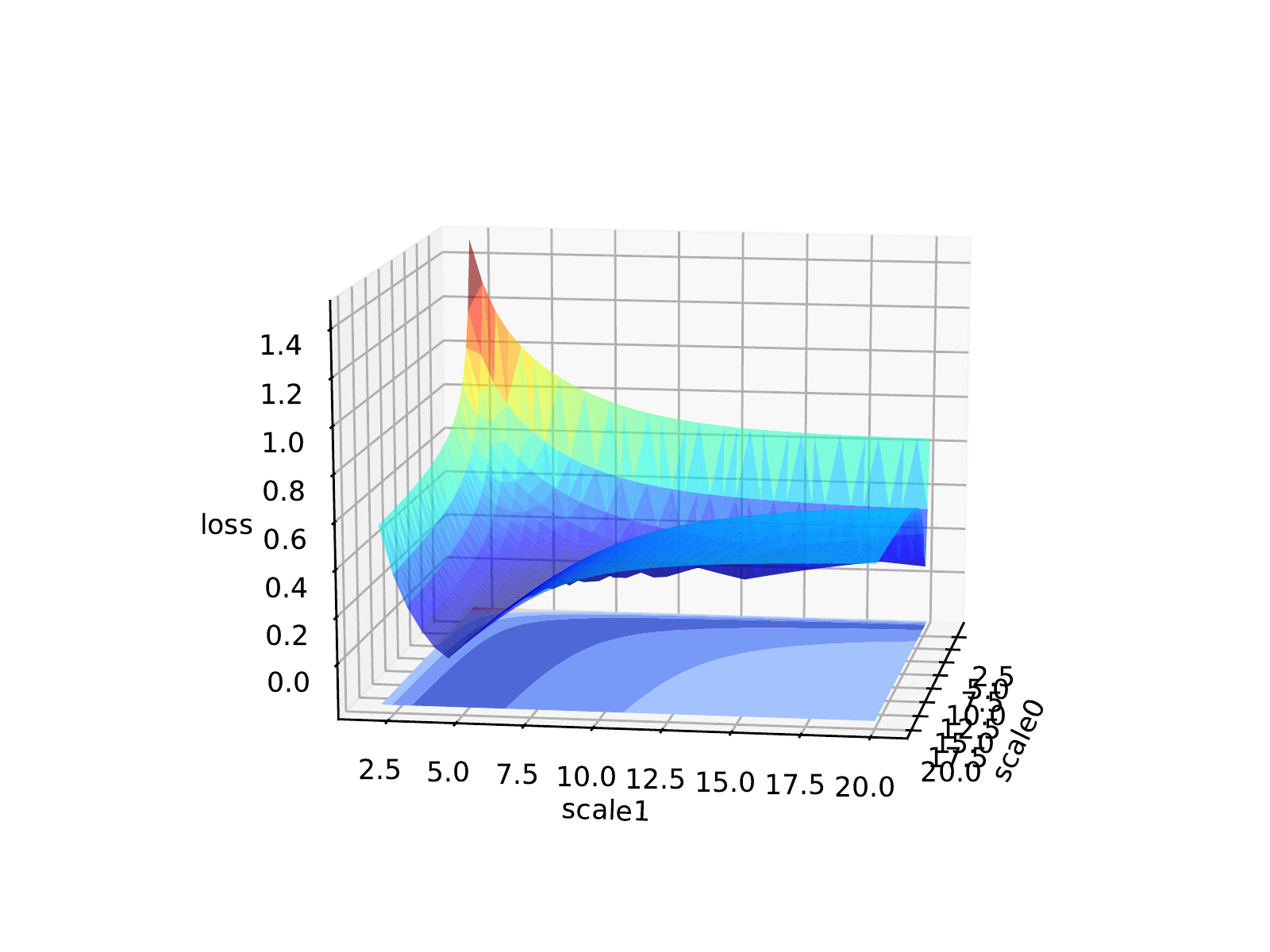}
    \end{center}
    \caption{\label{fig:aboveT-opt} Optimization objective for $\gamma(\mathsf{aboveT^\bullet})$}
\end{figure*}

\paragraph{Computing the objective efficiently.}
Each round in the evolutionary search requires computing the objective for each
candidate noise vector $\cvec_i$ in the population---the objective depends on
the probabilities of output events $\conc(\fsk{\cvec_i}(d)) \in E$, where
$\conc(\fsk{\cvec_i}(d))$ is the sketch $\sk$ with concrete noise scales~$\cvec_i$, applied to the input arguments from $\gamma$ and database~$d$. The
main new difficulty in our setting, compared to work on prior
testing/verification of differential
privacy~\citep{bichsel2018dp,ding2018detecting,albarghouthi2017synthesizing,zhang2016autopriv},
is that the noise scales are not fixed in advance: the search considers a population
of mechanisms with concrete different noise scales, and the scale parameters are
repeatedly adjusted as the optimization unfolds.

Since computing the fitness function is the primary bottleneck for an
evolutionary algorithm, it is crucial to efficiently compute the probabilities
of output events for different mechanisms. While there are existing symbolic
methods for computing probabilities exactly, these methods are too slow for our
purpose. Instead, a natural idea is to estimate output probabilities by
repeated sampling. This is more efficient than exact computation, but there are
still difficulties: estimating probabilities separately for each candidate
$\cvec_i$ introduces a large overhead when this procedure is employed on large 
populations across many optimization steps, and
stochastic variations across sampling steps due to random sampling introduce
more instability into the optimization procedure, leading to slower convergence.

To speed up this process, we take advantage of the structure of the objective.
For every example $\tuple{d_1, d_2, E} \in \exs$, computing the privacy loss $L$
for a candidate mechanism $M$ amounts to computing the probability of $E$ on
$d_1$, computing the probability of $E$ on $d_2$, and then taking the ratio of
the probabilities. Both of these steps can be optimized:
\begin{itemize}[leftmargin=*]
  \item
    \textbf{Computing the probability of an output event.}
    Instead of drawing random samples separately for every mechanism in the
    population, we can preselect a single set of random samples for each
    sampling statement in the program, and reuse these samples---across all
    candidates and over all optimization steps---to estimate the probability of the output event $E$. This leads to a significant reduction in the sampling time. 
        Of course,
    since a specific noise value may have different probabilities under the 
    different candidates, we must weight each trial differently for each
    candidate when estimating each probability of $E$.

    In more detail, suppose the sketch $\sk$ has $n$ holes. We draw $m$ uniform
    vectors $\mathbf{v}_j \in \mathbb{R}_+^n$ of non-negative real numbers
    representing the \emph{results} of the sampling statements in $\sk$. Then,
    each probability of $[\conc(\fsk{\cvec_i}(d)) \in E]$ can be estimated by
    counting how many runs of~$\conc(\sk)(d)$ with noise $\mathbf{v}_j$ produce
    an output in $E$, weighted by the probability of drawing $\mathbf{v}_j$ from
    the Laplace distribution with noise scale $\cvec_i$; this probability can be
    computed analytically and efficiently, without sampling. The latter step is
    essentially an application of Monte Carlo integration; to reduce variance
    further, our implementation performs importance sampling using a fixed
    Laplace (or Exponential) distribution as the proposal distribution. 
  \item
    \textbf{Computing the privacy loss at an output event.}
    After computing the probability of $E$ on inputs $d_1$ and $d_2$, we must
    take the ratio of these probabilities to compute the privacy loss, and then
    take a maximum over all examples. To further reduce the number of samples
    required, we can estimate the probability on $d_1$ and $d_2$ using the same
    set of samples; this correlated sampling method was previously used
    by~\citet{bichsel2018dp}.
\end{itemize}


\paragraph{Running example: Above Threshold.}
To give a better idea of the optimization problem for our running example,
\cref{fig:aboveT-opt} shows the objective (sans regularization) as we plug in
different concrete noise values for the holes in $\mathsf{aboveT}^\bullet$, with
arguments fixed to $\gamma$. To make the space easier to visualize, the
plot only varies noise for the first two locations while $\eta_3$ is set to $\bot$. 
The plot shows that there
isn't a single noise setting that minimizes the objective; rather, there is a
broad \emph{region} where the objective is approximately minimized (visualized
in the contour map in dark blue).

\begin{figure*}[h]
    \begin{subfigure}{0.45\textwidth}
\centering
        \includegraphics[scale=0.5]{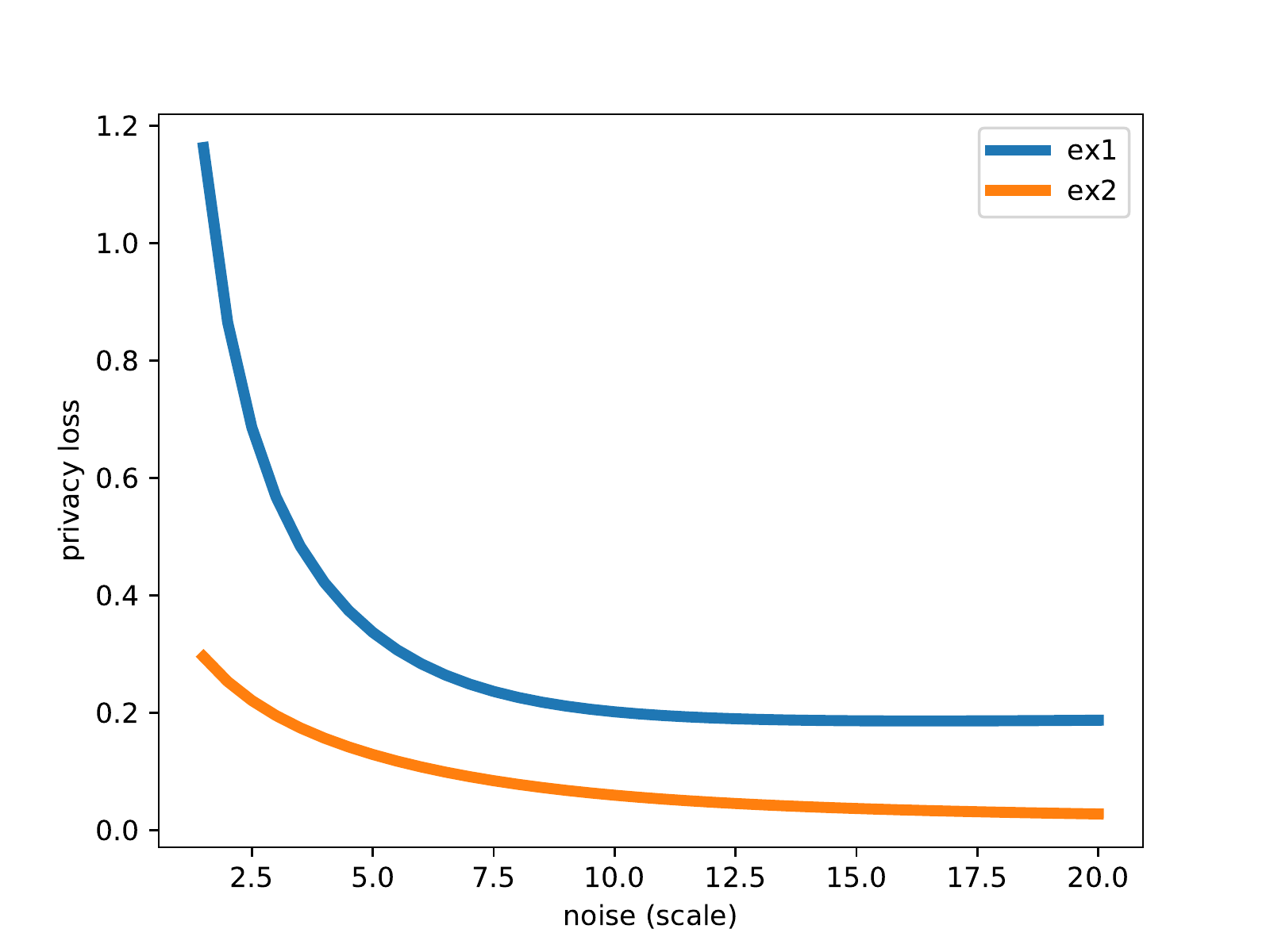}
        \caption{\label{fig:noisy_max_distinguish} Privacy loss for different examples}
    \end{subfigure}%
    \hfill
    \begin{subfigure}{0.45\textwidth}
\centering
    \includegraphics[scale=0.5]{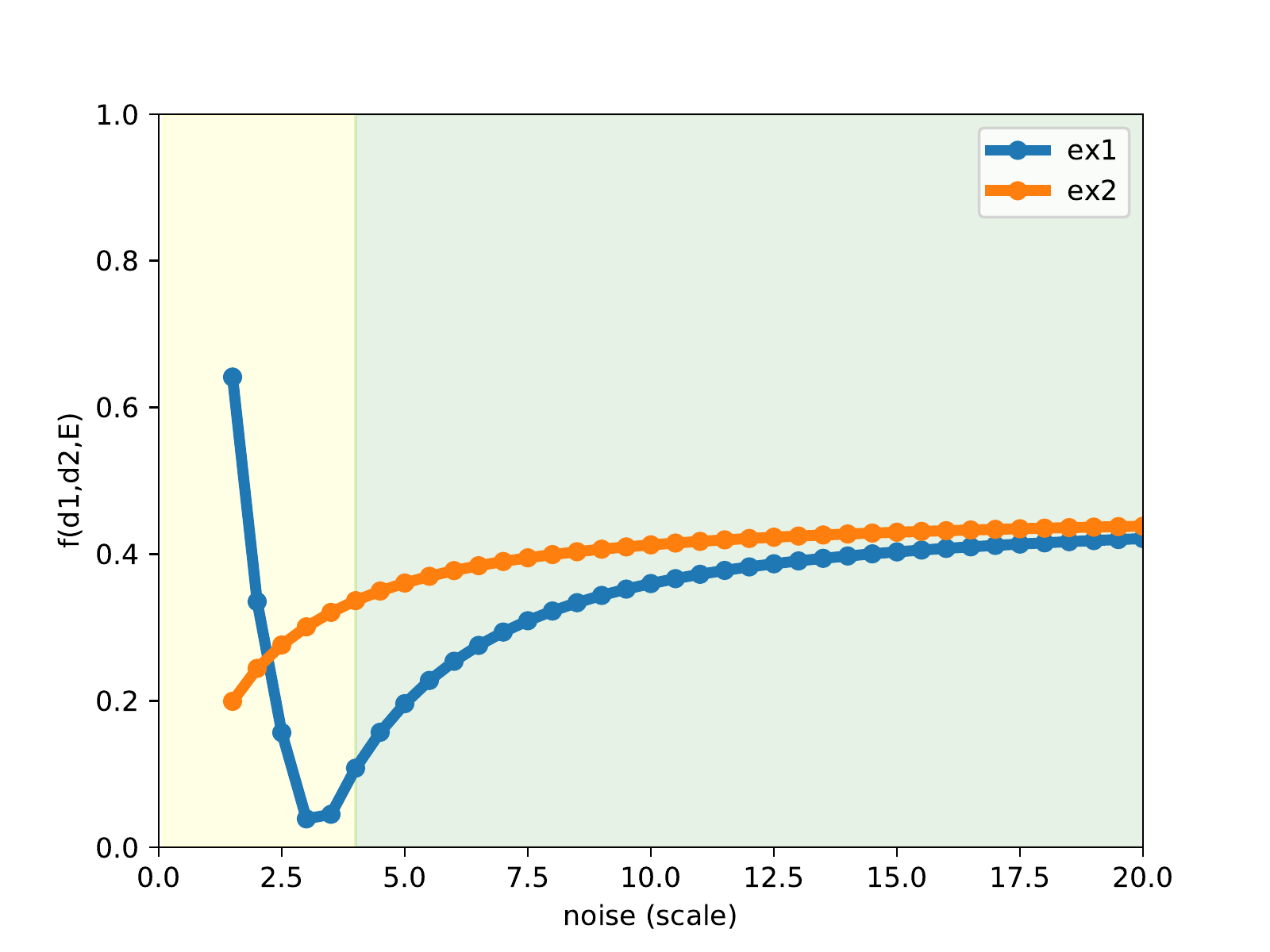}
        \caption{\label{fig:noisy_max_objective} $f(d_1, d_2, E)$ at $\conc{(\epsilon)}=0.5$; the green shade captures the region where this algorithm is 0.5-differentially private}
    \end{subfigure}
    \caption{Comparing the implication of the choice of examples}
\end{figure*}

\begin{figure}
    \[
      \begin{array}{l}
        \mathsf{NoisyMax}^\bullet(d, \vec{q}, \epsilon): \\
        \quad m, v \gets \bot, \bot; \\
        \quad \vec{a} \gets \vec{q}(d);  \\
        \quad \mathsf{for}\ i \in \{1\dots |q|\} \\
        \quad\quad b \gets \vec{a}_i + {\Lap}({\boxed{\eta}}); \\ 
        \quad\quad \mathsf{if}\ b \geq v\\
        \quad\quad\quad m, v \gets i, b; \\
        \quad \mathsf{return}(m);
      \end{array}
    \]


  \caption{NoisyMax1: Sketch program}
  \label{fig:ex-noisymax-sketch}
\end{figure}

\subsection{Enumerative synthesis}
\label{sec:enumsyn}
To complete synthesis, we want to produce symbolic expressions that lead to
noise close to the noise region $R$; these completions should work for various
settings of the mechanism arguments, not just $\conc$.  Our procedure to search
for an expression, denoted $\search$ in \cref{fig:synth}, inspired by the
classical synthesize-check loop from the formal-methods literature, enumerates
all expressions $\evec$ in the grammar $G$. However, it only considers
expressions where the concrete values $\conc(\evec)$ are in a
\emph{neighborhood} of the noise region $R$, denoted $\nbhd(R)$ (defined on the
$L_1$ distance of instantiations in $R$). This pruning step is the key to
efficient synthesis (see \cref{sec:eval} for a more thorough evaluation).
Formally, we generate the following set of candidate expressions:
\[
  \mathit{cand} = \{ \evec \mid \evec \in G, \; \conc(\evec) \in \nbhd(R)\}
\]
In general, this set contains multiple candidate expressions. To narrow down
this list, we augment $\exs$ with additional test examples
$\exs_\emph{test}$---varying the arguments $\conc$---and then rank the
candidates according to (1) how many examples in $\exs_\emph{test}$ they violate
(fewer is better), (2) their privacy loss on examples in $\exs_\emph{test}$
(higher is better), and (3) the magnitude of noise injected (smaller is better).
Note that we may accept expressions that violate some small number of examples,
since there is probabilistic noise involved when checking whether an example is
violated. Finally, we call the testing tool on the mechanisms from the top few
completions in $\mathit{cand}$, and output a ranked list of passing mechanisms 
(following the same ranking as above) as candidate solutions to the mechanism synthesis problem.

\paragraph{Importance of selecting challenging examples.}
A key aspect of our approach is selecting ``challenging'' examples. To better
understand why certain examples may be easier or more difficult than others, let
us consider the \textit{Noisy Max} mechanism~(\cref{fig:ex-noisymax-sketch}): it returns the index of an approximately
maximum query while preserving privacy by perturbing every query by Laplace
distributed noise. It is known that this mechanism achieves
$\epsilon$-differential privacy when the noise scale $\eta$ is taken to be~$2/\epsilon$~\citep{DR14}.

\Cref{fig:noisy_max_distinguish} plots the privacy loss (vertical axis) of this
mechanism for different concrete values of $\eta$ (horizontal axis) for two
different examples $\langle d_1, d_2, E\rangle$, {$ex_1:\langle 111, 022,
1\rangle$} and {$ex_2:\langle 111, 022, 2\rangle$}. The plot shows that $ex_1$
is a more \textit{challenging example}: it incurs a higher privacy loss than
$ex_2$ for all values of $\eta$; so any expression synthesized to satisfy $ex_1$
will also satisfy $ex_2$. Conversely, symbolic expressions synthesized to ensure
differential privacy at $ex_2$ may not be capable of ensuring differential
privacy at $ex_1$. When deciding which examples to use, our synthesis procedure
should seek to keep examples like $ex_1$, and discard examples like $ex_2$---the
former example is \emph{more useful} than the latter example.

However, given two examples, it is not easy to tell which (if either) example is
more challenging. The objective of our optimization function can be viewed as a
rough heuristic, preferring more useful examples. To see this, consider the
following function that captures the behavior of our objective function on a
given example $\tuple{d_1, d_2, E}$:
\[
  f(d_1, d_2, E)(\cvec) = \left| \log(\loss(\conc(\fsk{\cvec}), d_1, d_2,\conc(\epsilon),E)) - {\conc(\epsilon)} \right|
\]
That is, for a provided example $\langle d_1, d_2, E\rangle$, $f(d_1, d_2, E)$
is a function that maps $\cvec$ to the value of the objective function when the
arguments to the noise distribution is $\cvec$. Roughly speaking, this function
measures how closely the mechanism satisfies $\conc({\epsilon})$-differential
privacy at the given example: noise scales $\cvec$ where the mechanism
$\fsk{\cvec}$ satisfies a much weaker or a much stronger guarantee have larger
objective, and the optimization process avoids these noise vectors $\cvec$.
While we do not have a rigorous proof that examples with lower objective are
more useful, we found this to be the case empirically.

For an example of this heuristic in action, the plot in
\cref{fig:noisy_max_objective} at $\conc(\epsilon)=0.5$ shows that the objective
function is lower at $ex_1$ than at $ex_2$ for all noise scales where the
mechanism does satisfy differential privacy. However, this plot also shows that
the example $ex_1$ is not the most challenging example possible. To see why this
is so, recall that \textit{Noisy Max} achieves $\epsilon$-differential
privacy when the scale parameter on the noisy distribution is $2/\epsilon$,
e.g., for $\epsilon=0.5$, the noise scale should be $4$. Thus, an example where
the privacy guarantee is tight should achieve $\epsilon$-differential privacy
only when the noise scale is $4$. The (blue) curve for $ex_1$ in
\Cref{fig:noisy_max_objective} achieves a minimum around noise scale $\eta
\approx 3$, so achieving $\epsilon$-differential privacy at example $ex_1$
requires less noise than is required for achieving $\epsilon$-differential
privacy for the whole mechanism (i.e., at all examples). While $ex_1$ is not
optimal in this sense, this example was sufficiently good to guide our
implementation to a successful synthesis of the optimal symbolic expression of
the noise parameter; we will describe our implementation in more detail in
\cref{sec:eval}.

\paragraph{Running example: Above Threshold.}
On this example, we use a grammar of expressions of the following form:
\[
  [-] \times |\vec{q}|^{[-]} \times T^{[-]} \times (1/\epsilon)^{[-]}
  \quad \text{or} \quad \bot
\]
where each unknown $[-]$ ranges over non-negative integers, and $\bot$
represents no noise added. In general, our expression grammar combines powers of
basic numeric expressions involving inputs of the target sketch.

For the sketch $\mathsf{aboveT}^\bullet$, our tool produces expression vectors
$(2/\epsilon, 4/\epsilon, \bot)$ and $(3/\epsilon, 3/\epsilon, \bot)$ as
candidate solutions. The first setting recovers the textbook version of the
Above Threshold algorithm~\citep{DR14}, while the latter gives a new variant
that is also differentially private. The privacy proof for the second variant
follows by adjusting parameters in existing privacy proofs of Above Threshold
(e.g.,~\citep{barthe2016proving}).

\section{Experiments}
\label{sec:eval}


{

\begin{figure}[t]
 \small  
\begin{minipage}[b]{0.49\textwidth}
    \[
      \begin{array}{l}
        \mathsf{Histogram}^\bullet(d, \vec{q}, \epsilon): \\
        \quad \vec{hist} \gets \vec{q}(d); \\ 
        \quad \vec{ans} \gets \vec{hist} + \overrightarrow{\Lap}(\boxed{\eta}); \\
        \quad \mathsf{return}~\vec{ans};
      \end{array}
    \]
    \captionof{figure}{Histogram: Sketch program}
  \label{fig:ex-histogram-sketch}
\end{minipage}

%
%

\end{figure}

\begin{figure*}[t]

 \small  
\begin{minipage}[t]{0.32\textwidth}
    \[
      \begin{array}{l}
        \mathsf{Sum}^\bullet(d, \vec{q}, \epsilon): \\
        \quad \vec{a} \gets \vec{q}(d);\\
        \quad s \gets 0;\\
        \quad \mathsf{for}\ i \in \{1\dots |\vec{q}|\}:\\
        \quad\quad s \gets s + a_i + {\Lap}(\boxed{\eta}); \\
        \quad \mathsf{return}~s;
      \end{array}
    \]
    \captionof{figure}{Sum: Sketch program}
  \label{fig:ex-sum-sketch}
\end{minipage}\hfill%
%
%
%
%
  \begin{minipage}[t]{0.32\textwidth}
    \[
      \begin{array}{l}
        \mathsf{NoisyMax2}^\bullet(d, \vec{q}, \epsilon): \\
        \quad m, v \gets \bot, \bot; \\
        \quad \vec{a} \gets \vec{q}(d);  \\
        \quad \mathsf{for}\ i \in \{1\dots |\vec{q}|\}: \\
        \quad\quad b \gets a_i + {\Lap}({\boxed{\eta_1}}); \\ 
        \quad\quad \mathsf{if}\ b > v\\
        \quad\quad\quad m, v \gets i, b; \\
        \quad ans \gets m + {\Lap}(\boxed{\eta_2}); \\
        \quad \mathsf{return}~ans;
      \end{array}
    \]
    \captionof{figure}{NoisyMax2: Sketch program}
  \label{fig:ex-noisymax2-sketch}
\end{minipage}\hfill%
%
%
  \begin{minipage}[t]{0.32\textwidth}
    \[
      \begin{array}{l}
        \mathsf{ExpNoisyMax}^\bullet(d, \vec{q}, \epsilon): \\
        \quad m, v \gets \bot, \bot; \\
        \quad \vec{a} \gets \vec{q}(d);  \\
        \quad \mathsf{for}\ i \in \{1\dots |\vec{q}|\}: \\
        \quad\quad b \gets a_i + {\Exp}({\boxed{\eta_1}}); \\ 
        \quad\quad \mathsf{if}\ b > v\\
        \quad\quad\quad m, v \gets i, b; \\
        \quad ans \gets m + {\Exp}(\boxed{\eta_2}); \\
        \quad \mathsf{return}~ans;
      \end{array}
    \]
    \captionof{figure}{ExpNoisyMax: Sketch program}
  \label{fig:ex-expNoisyMax-sketch}
\end{minipage}
\end{figure*}

%
%
%
%
%

%
%

\begin{figure*}
 \small  
  \begin{minipage}{0.32\textwidth}
  \[
    \begin{array}{l}
      \mathsf{AboveT1}^\bullet(d, \vec{q}, T, \epsilon): \\
      \quad i \gets 1; \\
      \quad done \gets \mathsf{false}; \\
      \quad t \gets T + \Lap(\boxed{\eta_1}); \\
      \quad \vec{a} \gets \vec{q}(d) + \overrightarrow{\Lap}(\boxed{\eta_2}); \\
      \quad \mathsf{while}\ i \leq |\vec{q}| \land \lnot done\ \mathsf{do} \\
      \quad\quad \mathsf{if}\ a_i > t\ \mathsf{then} \\
      \quad\quad\quad done \gets \mathsf{true}; \\
      \quad\quad i \gets i + 1; \\
      \quad\quad \mathsf{if}\ done\ \mathsf{then} \\
      \quad\quad\quad ans \gets i - 1; \\
      \quad\quad \mathsf{else} \\
      \quad\quad\quad ans \gets 0; \\
      \quad \mathsf{return}~ans;
    \end{array}
  \]
\caption{AboveT1: Sketch program}
\label{fig:ex-aboveT1-sketch}
\end{minipage}\hfill%
  \begin{minipage}{0.32\textwidth}
    \[
      \begin{array}{l}
        \mathsf{SVT}^\bullet(d, \vec{q}, N, T, \epsilon): \\
           \quad    out \gets [\ ]; i \gets 1; \\
           \quad    count \gets 0; \\
           \quad    t \gets T + {\Lap}(\boxed{\eta_1}); \\
           \quad    \vec{a} \gets \vec{q}(d); \\
           \quad    \mathsf{while}\ i \leq |\vec{q}|: \\
           \quad\quad        qans = a_i + {\Lap}(\boxed{\eta_2}); \\
           \quad\quad        \mathsf{if}\ qans > t; \\
           \quad\quad\quad            \mathsf{append}(out, \top); \\
           \quad\quad\quad            count \gets count + 1; \\
           \quad\quad\quad            \mathsf{if}\ count \geq N\ \mathsf{then} \\
           \quad\quad\quad\quad                \mathsf{break}; \\
           \quad\quad        \mathsf{else} \\
           \quad\quad\quad            \mathsf{append}(out, \bot); \\
           \quad\quad i \gets i + 1; \\
           \quad    \mathsf{return}\ out; 
      \end{array}
    \]
  \caption{SVT: Sketch program}
  \label{fig:ex-svt-sketch}
\end{minipage}\hfill%
%
%
  \begin{minipage}{0.32\textwidth}
    \[
      \begin{array}{l}
        \mathsf{SmartSum}^\bullet(d, \vec{q}, M, \epsilon): \\
   \quad n \gets 0; i \gets 1; \\
   \quad next \gets 0; sum \gets 0; \\
   \quad \vec{a} \gets \vec{q}(d); \\
   \quad r \gets [\ ]; \\
   \quad \mathsf{while}\ i \leq \left | \vec{q} \right |: \\
   \quad\quad     sum \gets sum + a_i; \\
   \quad\quad     \mathsf{if}\ i\ \mathsf{mod}\ M = 0\ \mathsf{then} \\ 
   \quad\quad\quad         n \gets n + sum + {\Lap}(\boxed{\eta_1}); \\
   \quad\quad\quad         sum \gets 0; \\
   \quad\quad\quad         next \gets n; \\
   \quad\quad     \mathsf{else} \\
   \quad\quad\quad         next \gets next + a_i + {\Lap}(\boxed{\eta_2}); \\
   \quad\quad     \mathsf{prepend}(r, next); \\
   \quad\quad     i \gets i + 1; \\
   \quad \mathsf{return}\ r; \\
      \end{array}
    \]
  \caption{SmartSum: Sketch program}
  \label{fig:ex-smartsum-sketch}
\end{minipage}
\end{figure*}

%
%
}

We developed \toolname, an implementation of our synthesis procedure. Our
implementation is written in the Julia programming language~\cite{Julia}, and
uses the \textit{BlackBoxOptim} package~\cite{bboptim} for optimization. 


\paragraph{Benchmark examples.}

We used our tool to synthesize nine mechanisms from the differential privacy
literature.  We briefly describe our benchmarks here. We chose this set of benchmarks because they
represent foundational algorithms and have been previously considered in the
formal verification and testing
literature~\cite{ding2018detecting,bichsel2018dp,albarghouthi2017synthesizing,wang2019proving,barthe2016proving}.

\begin{description}[itemsep=0pt, topsep=0pt]
    \item[Sum] One of the simplest benchmarks,
      \textit{Sum}~(\cref{fig:ex-sum-sketch}) computes the sum of a private
      list of numbers. To ensure privacy, each list element is perturbed by
      noise drawn from a Laplace distribution.
    \item[Histogram] Given a input database,
      \textit{Histogram}~(\cref{fig:ex-histogram-sketch}) computes a histogram
      of frequency counts of number of elements in predefined buckets. It
      ensures privacy by adding noise from a Laplace distribution to each of the
      frequency counts. 
    \item[Above Threshold] The benchmark
      \emph{AboveT1}~(\cref{fig:ex-aboveT1-sketch}) is a sketch of the
      AboveThreshold algorithm~\citep{DR14}. This program takes a numeric
      threshold and a list of numeric queries, and returns the index of the
      first query whose answer is (approximately) above the threshold. We also
      consider a variant \emph{AboveT2}~(\cref{fig:ex-aboveT-sketch}), where the
      sketch has more noise locations than required.
    \item[Sparse Vector Technique] The benchmark
      \emph{SVT}~(\cref{fig:ex-svt-sketch}) is a sketch of the Sparse Vector
      Technique, an algorithm that has been rediscovered numerous times in the
        differential privacy literature~\citep{DR14}. The variant of SVT that we use~\citep{Lyu2017} returns a vector to indicate which of the queries are above/below a noisy threshold; the mechanism halts after it outputs $N$ above noisy threshold ($\top$) responses. 
    \item[Noisy Max] The benchmark \emph{NoisyMax1}~(\cref{fig:ex-noisymax-sketch})
      is a sketch of the Report-Noisy-Argmax algorithm from the differential privacy
      literature~\citep{DR14}, which takes a list of numeric queries and releases
      the index of the query with (approximately) the highest answer. The sketch for
      \emph{NoisyMax2}~(\cref{fig:ex-noisymax2-sketch}) has more locations than
      required, and we also consider a variant
      \emph{ExpNoisyMax}~(\cref{fig:ex-expNoisyMax-sketch}) where the sketch
      specifies noise drawn from the Exponential distribution instead of the Laplace
      distribution.
    \item[SmartSum] The benchmark
      \emph{SmartSum}~(\cref{fig:ex-smartsum-sketch}) implements the two-level
      counter mechanism for computing all running sums of a
      sequence~\citep{CSS10,DNPR10}; roughly speaking, it chunks the sequence
      into blocks and adds noise to each block. The algorithm requires addition
      of noise at two program locations.
\end{description}

\paragraph{Comparison with simpler procedures.}
Our synthesis method involves quite a few moving parts. To demonstrate the
importance of each phase in our algorithm, we evaluate \toolname against four
simpler baselines. In order of increasing sophistication:
\begin{description}[itemsep=0pt, topsep=0pt]
  \item[na\"ive] The na\"ive baseline applies a brute-force strategy: it
    enumerates all expressions at all program locations, queries the tester
    \statdp~\cite{ding2018detecting} as an oracle to accept or reject each
    choice of expressions, and then finally ranks all combinations of
    expressions. This baseline is perhaps the simplest synthesis method.
  \item[unlim] This baseline improves upon brute-force enumeration by including
    an unbounded \textit{counterexample cache} to memoize ``good''
    counterexamples that have been able to cause violations for past instances.
    The idea behind this baseline is the hypothesis that a few ``good''
    counterexamples can cause violations for most of the candidates being
    enumerated, so there is no need to find fresh counterexamples for every
    candidate. The cache is sorted by (1) \textit{utility} (the number of
    expression vectors that created a violation for this counterexample), and
    (2) \textit{recency} (how new the counterexample is). Candidate mechanisms
    are evaluated, in order, on the examples in the counterexample cache before
    spawning the expensive tester \statdp to discover a new counterexample; if
    \statdp can produce a new counterexample, it is added to the cache.
  \item[lim] When the counterexample cache grows too large, the \emph{unlim}
    baseline wastes time searching through the cache on ``bad'' counterexamples
    at the tail end of the cache. In this optimized version, we limit the
    counterexample cache to the top-5 counterexamples (we empirically found this
    size to be a good setting).
  \item[noopt] This version operates similarly to \toolname but it does not use
    optimization to find a region $R$ of noise values: all expressions that
    satisfy the grammar of expressions~($G$) are ranked according to the
    heuristics described in~\cref{sec:enumsyn}, and the top-k ranked expressions~\mbox{(k=5$\times$ \#locations)} are sent for verification. Compared to the other
    baselines, this version includes the initialization phase of \toolname where
    challenging examples are selected, and it uses a top-5 limited
    counterexample cache.
\end{description}

We ran our experiments on a cluster with 4 worker threads, provisioning for 4
cores and 8 GB memory for each task. \Cref{tab:results} shows our comparisons
with the baseline: the second column (\#locs) shows the number of locations in
the sketch where noise can be added. The blank cells in the experimental results
(\cref{tab:results}) correspond to jobs that did not complete on the cluster
over two days. 

\begin{table*}[!ht]

\begin{center}
    \normalsize
    \caption{\label{tab:results} \toolname versus baselines (time rounded to the nearest minute)}
    \rowcolors{3}{}{gray!10}

    \begin{tabular}{ lrrrrrrrrrrr}
 \toprule
    \multicolumn{2}{c}{Mechanism} & \multicolumn{6}{c}{\toolname (time in minutes and rank)} & \multicolumn{4}{c}{Baselines (time in minutes)} \\ 
    \cmidrule(lr){1-2} \cmidrule(lr){3-8} \cmidrule(lr){9-12}
    benchmark &  \#locs &  init &  opti &  enum &  verify &  \textbf{total} &  rank &  na\"ive & unlim & lim & noopt\\
 \midrule
    Histogram &  1 &  4 &  1 &  < 1 &  15 &  \textbf{20} &  1 &  55 & 98  &  74 &  60 \\  
    Sum &  1 &  3 &  < 1 &  < 1 &  10 &  \textbf{15} &  1 & 55  &  69 &  68 &  52 \\
    NoisyMax1 &  1 &  9 &  1 &  1 &  20 &  \textbf{32} &  1 & 84  &  63 & 61 &  54 \\
    NoisyMax2 &  2 &  14 &  2 &  6 &  48 &  \textbf{72} &  1 & - & - & - & - \\
    ExpNoisyMax &  2 &  32 &  1 &  7 &  50 &  \textbf{90} &  1 & - & - & - &  217 \\
    SmartSum &  2 &  31 &  6 &  12 &  42 &  \textbf{91} & 4  & - & - & - &  289\\
    SVT &  2 &  18 &  2 &  4 &  18 &  \textbf{44} &  1 & - & - & - & 128\\
    AboveT1 &  2 &  13 &  2 &  6 &  38 &  \textbf{60} & 1  & 1582  & - &  1647 &  115 \\  
    AboveT2 &  3 &  24 &  7 &  41 &  54 &  \textbf{126} &  2 & - & - & - & - \\  


 \bottomrule
\end{tabular}
\end{center}
\end{table*}


We briefly comment on the performance of the baseline solutions. The
\textit{na\"ive} baseline performs reasonably well for mechanisms that require
noise at a few locations with simple noise expressions (like \textit{Histogram}
and \textit{Sum}); however, it struggles when the number of locations and the
complexity of the expressions increase.

The baselines with the counterexample cache (\textit{lim} and \textit{unlim})
have improved performance in some cases (e.g., \textit{NoisyMax1}). For the
simpler cases (\textit{Sum} and \textit{Histogram}), it seems that the baseline
loses too much time in warming up the cache. However, for a slightly more
involved benchmark (\textit{NoisyMax1}), the cache pays off. Also, in general,
limiting the cache size~(\textit{lim}) seems to be a better configuration than
an unlimited cache~(\textit{unlim}). The behavior of \textit{AboveT} is a bit
surprising: it fails to complete for \textit{unlim}, which it understandable,
but draws similar runtimes for both \textit{na\"ive} and \textit{lim}; it seems
that the overhead of using the cache cancels out its gains.

The performance of the \textit{noopt} baseline is a significant improvement over
the previous baselines: it synthesizes most of the mechanisms in reasonable time
and solves many more benchmarks, especially, the more involved ones. This shows
the importance of identifying ``good'' counterexamples. 

Finally, \toolname, by including the optimization phase, is~2$\times$ to
20$\times$ faster than \textit{noopt}. Furthermore, it solves a couple of
benchmarks that \textit{noopt} could not complete. This shows the value of the
optimization phase, especially for involved mechanisms with more noise
locations.

Overall, the above experiment shows that selecting challenging counterexamples
and identifying a noise region for the search for expressions are crucial to the
success of \toolname.

\paragraph{Time spent in different phases.}
\toolname spends most of its time in the tester \statdp, either in search of
representative counterexamples (\textit{init}) or in the final verification
(\textit{verify}). The optimization phase (\textit{opti}) and enumerative
synthesis (\textit{enum}) are reasonably fast. The optimization phase owes its
speed to our technique of approximating the probability estimates using
importance sampling and the heuristic of reusing noise samples across candidates.
The total time (\textit{total}) is the end-to-end time taken by \toolname,
including time for logging.

\paragraph{Ranking of mechanisms.}
For almost all benchmarks, the noise setting corresponding to the textbook
versions of these mechanisms is discovered by our tool and ranked high. Recall
that we run a continuous optimization to prepare an preliminary ranking of the
candidates, and the top ranked examples are passed to \statdp for a more
rigorous test for the final ranking.

For each of \emph{Sum}, \emph{Histogram} and \emph{NoisyMax1}, the textbook mechanism
is ranked among the top two candidates even after the preliminary ranking, which, then,
emerges as the topmost candidate after the final ranking. This
behavior does not change for \emph{NoisyMax2} where we add an additional noise
location in the sketch, showing that our method can ignore irrelevant noise
locations---when sketches are obtained by annotating existing, non-private code,
many of the possible noise locations may be unnecessary for privacy. Our tool
performed similarly for \emph{ExpNoisyMax}, which not only contains an
additional noise location but also uses an Exponential noise distribution
instead of the Laplace distribution, showing that our method can be applied to
sketches with noise distributions besides Laplace.

For \emph{AboveT1}, the top two solutions that emerged are $(2/\epsilon,
4/\epsilon)$ and $(3/\epsilon, 3/\epsilon)$. The former is the classic version
of the algorithm~\citep{DR14}, while the latter is a \textit{new} variant
identified by \toolname. The proof of privacy of the new variant follows the proof for the standard
variant (see, e.g., \citet{barthe2016proving}). In fact, the existing privacy
proof applies to \emph{exactly} these two variants, and no other variants. This
benchmark shows that our tool is able to automatically discover new versions of
well-known private algorithms.

For \emph{SVT}, the standard version of the mechanism was not the topmost in the
preliminary ranking, though it was high enough to be selected for the final
phase, where it was identified it as the topmost candidate.

The only benchmark in which our tool had difficulty was \emph{SmartSum}, where the
ideal solution ended up being ranked fourth in the final ranking. While the
expected noise expressions are $(2/\epsilon, 2/\epsilon)$, our procedure
proposed the mechanism with noise scale $(1/\epsilon, 2/\epsilon)$ as the top
ranking candidate because \statdp could not find counterexamples against the
mechanism with $(1/\epsilon, 2/\epsilon)$, even though this mechanism is not
$\epsilon$-differentially private. Thus, \statdp did not generate high-quality
examples (in $\selex$) to direct the search away from the incorrect expression
and towards the correct expressions. In most of our benchmarks, however, we
found that \statdp performed quite well.

\section{Related Work}
\label{sec:rw}

\paragraph{Program synthesis.}
Program synthesis is an active area of research; we summarize the most related
directions here. Closest to our work is the recent paper 
\citet{smith2019synthesizing} that develops a technique relying on user-defined
examples to synthesize private programs in a strongly-typed functional language.
However, this approach can only synthesize simple mechanisms where the privacy
analysis follows from standard composition theorems; even if provided with an
infinite number of examples, their system is not be able to synthesize
mechanisms like NoisyMax, SVT, AboveT, and SmartSum. Our synthesis technique is
also radically different: rather than using a type-directed approach, we perform
a reduction to continuous optimization.

In terms of synthesizing randomized algorithms generally, most work has focused
on programs where all inputs are known; in that setting, the target
specification for the synthesis problem is simpler---there is no need to
quantify over all inputs, unlike the universal quantification over pairs of
databases in our
setting~\citep{albarghouthi2017repairing,chaudhuri2014bridging}. Our general
approach of using a small number of examples to guide the search appears in
various forms of \emph{counterexample-guided inductive synthesis}
techniques~\citep{chaudhuri2014bridging}

\paragraph{Finding counterexamples to DP.}
There have been a few proposals for finding violations to differential privacy.
As mentioned, our approach builds heavily on \statdp~\citet{ding2018detecting}, a counterexample generation tool for differential privacy using statistical tests.

A different approach, DP-Finder~\citep{bichsel2018dp}, reduces the search
for counterexamples to an optimization problem by approximating the mechanism by
a differential surrogate function, thereby allowing the use of numerical
optimization methods. The solution of the optimization on the surrogate function
is a candidate counterexample. An exact solver (eg. PSI~\citep{gehr2016psi}), or an
approximate, sampling-based estimator is then used to check if the candidate is
a true counterexample on the actual mechanism. In spirit, our use of presampling
is similar to derandomization of candidates in DP-Finder. Our approach also
relies on an optimization problem but instead of transforming the optimization
space via a surrogate function, we first concretize the mechanism to transfer
the search over symbolic expressions to a search over real vectors, and then use
a black-box optimizer that does not require gradients. 

\textsc{CheckDP}~\citet{wang2020check} combines verification and falsification of differential privacy. Unlike \statdp,
\textsc{CheckDP} relies on symbolic, rather than statistical methods to prove
privacy and generate counterexamples. As a result, counterexamples do not come
with a $p$-value or measure of tightness, a measure that is crucial to our
example-selection process. It would be very interesting to see if the
counterexamples and more powerful analysis afforded by \textsc{CheckDP} could be
used to drive a synthesis approach, like ours.

\paragraph{Verifying DP.}
Differential privacy has been a prime target for formal verification ever since
it was introduced by Dwork et al.~\citet{DMNS06}, due to its compelling motivation, rigorous
foundations, and clean composition properties. There are too many verification
approaches to survey here, applying techniques like runtime verification,
various kinds of type systems, and program logics. Unlike our approach, all of
these approaches assume that the mechanism to be verified is fully specified.
The most advanced examples considered in our benchmarks (e.g., \emph{NoisyMax},
\emph{AboveT}, \emph{SVT}) have only recently been verified
\citep{barthe2016proving,albarghouthi2017synthesizing}; they have also been
tricky for human experts to design correctly~\citep{lyu2016understanding}.
LightDP~\citep{zhang2016autopriv} proposes a language for verifying
privacy-preserving mechanisms and dependent type system for annotations to
synthesize proofs of differential privacy. It constructs proofs by
\emph{randomness alignment} via an \emph{alignment function} that ``aligns" the
noise on the executions corresponding to the adjacent databases.
ShadowDP~\citep{wang2019proving} attempts to construct a randomness alignment by
instrumenting \emph{shadow executions} to transform a probabilistic program to a
program where privacy costs appear explicitly. This allows the transformed
program to be verified by off-the-shelf verification tools.

\section{Discussion \label{sec:discuss}}



We propose the first technique for automatically synthesizing complex differential privacy mechanisms (like NoisyMax, SVT, AboveT, and SmartSum) from sketches. Our approach does have certain limitations which opens up opportunities for interesting future work.

Perhaps the primary limitation of \toolname is its dependence on \statdp for
challenging counterexamples; developing techniques to generate high-quality,
``worst-case'' counterexamples would likely improve the synthesis procedure.

In the absence of a fast and robust verifier for differential privacy, we used a
testing tool as a stand-in for a verification oracle. We assume that the failure
to reject the null hypothesis (at a $\alpha$ = 0.05) is an indication that the
algorithm is DP at the provided privacy budget. This is a clear limitation of
our choice of using a tester as a verifier, given that testers and verifiers
answer complimentary questions: while a verifier ensures soundness (that a
verification instance that is claimed to be verified is indeed so), a tester, on
the other hand, guarantees completeness (any counterexample generated does
indicate a violation). Nevertheless, we found it to be a good choice in
practice. Once the candidates are ranked by \toolname, an existing differential
privacy verifier can then be used as the final step to prove that the
synthesized program is private; besides the new variant of Above Threshold, the
target mechanisms in our examples have all been certified by existing automatic
verifiers~\citep{albarghouthi2017synthesizing,zhang2016autopriv}.


Our algorithm requires, as inputs, a sketch of a mechanism with noise
expressions as holes and a finite grammar $G$ for noise expressions. The
algorithm is not capable of performing any syntactic transformations over the
input sketch. It will fail to find a solution if no such noise expressions exist
for the provided sketch within the provided grammar.


All our benchmarks were run with the same heuristics and the same setting of the
hyperparameters (see Appendix); 
however, synthesizing more mechanisms
would give a better assessment of the generality of these heuristics. Finally,
it would be interesting to consider other forms of privacy, e.g., $(\epsilon,
\delta)$-DP and R\'enyi differential privacy~\citep{BunS2016,MironovRDP}.


\paragraph{Acknowledgements.}
The first author is grateful to the United States-India Educational Foundation
for their support. This work is partially supported by the NSF (CNS-2023222,
CCF-1943130, CCF-1652140), and grants from Facebook.

\bibliography{header,biblio}
\bibliographystyle{IEEEtran}

\appendix

%

\section{Hyperparameters\label{sec:appendix}}

{\small
We use the following setting of hyperparameters for \textit{all} the mechanisms we evaluated (i.e. we did not tune them separately for each case). The hyperparameters were selected via a set of preliminary experiments on a few mechanisms. These values continued to hold well as our benchmark set was expanded with more mechanisms. Nevertheless, a more exhaustive study can be done to evaluate the generality of this setting.

\begin{itemize}[leftmargin=*]

    \item[] \textbf{Selecting Examples}
\begin{itemize}
    \item We pick the zone of confusion on \textit{p-values} $ \in [0.05,0.9]$. 
\end{itemize}

    \item[] \textbf{Region Selection}
\begin{itemize}
    \item For importance sampling, we use a distribution of the same family (Laplace or Exponential) as provided in the sketch and a scale of 4.0; 
    \item We use $\lambda=1$ for the regularization parameter in the objective function in the optimization phase i.e. we weigh each of the objective function and the simplicity of the expression equally.
\end{itemize}

    \item[] \textbf{Differential Evolution}
\begin{itemize}
    \item We ran our optimizer for (500$\times$\#locations) steps, where \#locations refers to the number of noise locations specified in the sketch; 
    \item The size of the population was set to 50. 
\end{itemize}

    \item[] \textbf{Enumerative Synthesis}
\begin{itemize}
    \item Our enumeration of expressions is over:
      \[ [1-4] \times |\vec{q}|^{[0-2]} \times (1/\epsilon)^{[1-2]} \quad \textrm{or} \quad \bot \]
        (noise expressions must be directly proportional to $\vec{q}$ and inversely proportional to $\epsilon$);
    \item We define the neighborhood (Nbhd) of the region $R$ as all instantiations lying within an L1 distance of 3 from instantiations in $R$;
    \item From the set of ranked expression vectors emitted by the enumerative synthesis phase, we select the top-(5$\times$\#locations) for rigorous verification.
\end{itemize}

\end{itemize}
}

\end{document}